\documentclass[12pt]{article}

\usepackage{scicite}

\usepackage{times}

\newenvironment{sciabstract}{%
\begin{quote} \bf}
{\end{quote}}

\newcommand{\ket}[1]{\vert#1\rangle}

\usepackage{amsmath}
\usepackage{txfonts}    	
\usepackage{color}		

\usepackage{graphicx}
\usepackage{caption}

\newcounter{lastnote}

\title{Universal digital quantum simulation with trapped ions}

\author{B. P. Lanyon$^{1,2\ast}$, C. Hempel$^{1,2}$, D. Nigg$^{2}$, M. M\"uller$^{1,3}$, R. Gerritsma$^{1,2}$, \\
F. Z\"ahringer$^{1,2}$, P. Schindler$^{2}$, J. T. Barreiro$^{2}$, M. Rambach$^{1,2}$, G. Kirchmair$^{1,2}$, \\
M. Hennrich$^{2}$, P. Zoller$^{1, 3}$, R. Blatt$^{1,2}$, C. F. Roos$^{1,2}$.\\
\\
\normalsize{$^{1}$Institut f\"ur Quantenoptik und Quanteninformation}\\
\normalsize{\"Osterreichische Akademie der Wissenschaften,}\\
\normalsize{Otto-Hittmair-Platz 1, A-6020 Innsbruck, Austria,}\\
\normalsize{$^{2}$Institut f\"ur Experimentalphysik, University of Innsbruck,} \\
\normalsize{Technikerstr. 25, A-6020 Innsbruck, Austria,}\\
\normalsize{$^{3}$Institut f\"ur Theoretische Physik, University of Innsbruck,} \\
\normalsize{Technikerstr. 25, A-6020 Innsbruck, Austria}.
\\
\normalsize{$^\ast$To whom correspondence should be addressed; E-mail:  ben.lanyon@uibk.ac.at}
}

\date{}

\begin{document} 


\baselineskip24pt


\maketitle


\begin{sciabstract}
A digital quantum simulator is an envisioned quantum device that can be programmed to efficiently simulate any other local system.  
We demonstrate and investigate the digital approach to quantum simulation in a system of trapped ions. Using sequences of up to 100 gates and 6 qubits, the full time dynamics of a range of spin systems are digitally simulated. Interactions beyond those naturally present in our simulator are accurately reproduced and quantitative bounds are provided for the overall simulation quality. 
Our results demonstrate the key principles of digital quantum simulation and provide evidence that the level of control required for a full-scale device is within reach. 
\end{sciabstract}


While many natural phenomena are accurately described by the laws of quantum mechanics, solving the associated equations to calculate properties of physical systems, i.e. simulating quantum physics, is in general thought to be very difficult~\cite{Fey82}. Both the number of parameters and differential equations that describe a quantum state and its dynamics, grow exponentially with the number of particles involved. One proposed solution is to build a highly controllable quantum system that can efficiently perform the simulations~\cite{Buluta02102009}. Recently, quantum simulations have been performed in several different systems~\cite{PhysRevLett.82.5381,Greiner:2002,Leibfried:2002,Friedenauer:2008qm,Gerritsma:2010,Kim:2010,Lanyon:2010,PhysRevLett.97.050504,Barreiro:2011,greiner2011}, largely following the analog approach~\cite{Buluta02102009} whereby an analogous model is built, with a direct mapping between the state and dynamics of the simulated system and those of the simulator. An analog simulator is dedicated to a particular problem, or class of problems. 

A digital quantum simulator~\cite{Llo96,Jane,Buluta02102009,wiebe} is a precisely controllable many-body quantum system on which a universal set of quantum operations (gates) can be performed, i.e. a quantum computer~\cite{NC01}. The simulated state is encoded in a register of quantum information carriers, and the dynamics are approximated with a stroboscopic sequence of quantum gates. What makes this device so special is that it can, in principle, be reprogrammed to efficiently simulate any local quantum system~\cite{Llo96} and is therefore referred to as a universal quantum simulator. Furthermore, there are known methods to efficiently correct for and quantitatively bound experimental error in large-scale digital simulations~\cite{error1}.
 
We report on digital simulations using a system of trapped ions. We focus on simulating the full time evolution of networks of interacting spin-1/2 particles, which are models of magnetism~\cite{Assa} and exhibit rich dynamics. We do not use error correction, which has been demonstrated separately in our system~\cite{Schindler27052011}, and must be included in a full-scale fault-tolerant digital quantum simulator.

The central goal of a quantum simulation is to calculate the time-evolved state of a quantum system $\psi (t)$.  
In the case of a time-independent Hamiltonian $H$ the form of the  solution is $\psi(t){=}e^{-i H t/\hbar}\psi(0){=}U\psi(0)$. A digital quantum simulator can solve this equation efficiently for any local quantum system~\cite{Llo96}, i.e. where $H$ contains a sum of terms $H_k$ that operate on a finite number of particles, due to interaction strengths that fall of with distance for example. 
In this case the local evolution operators $U_k=e^{-iH_kt/\hbar}$ can be approximated using a fixed number of operations from a universal set. However, these terms do not generally commute $U\neq\prod_{k}e^{-iH_kt/\hbar}$. This can be overcome using the Trotter approximation~\cite{Trotter,Llo96}, $e^{-iHt}=\lim_{n \to \infty} (\prod_ke^{-\frac{i}{\hbar}H_kt/n})^n$, for integer $n$, which is at the heart of the digital quantum simulation algorithm. For finite $n$ the Trotter error is bounded and can be made arbitrarily small. The global evolution of a quantum system can therefore be approximated by a stroboscopic sequence of many small time-steps of evolution due to the local interactions present in the system. The digital algorithm can also be applied to time-dependent Hamiltonians and open quantum systems~\cite{Llo96,NC01,PhysRevA.65.010101,wiebe}. 

Our simulator is based on a string of electrically trapped and laser-cooled calcium ions (see \cite{online}). 
The $\ket{S_{1/2}}{=}\ket{1}$ and $\ket{D_{5/2}}{=}\ket{0}$ Zeeman states encode a  qubit in each ion. Simulated states are encoded in these qubits and manipulated by laser pulses that implement the operation set: $O_1(\theta,j)=\exp(-i\theta\sigma^{j}_{z})$; $O_2(\theta)=\exp(-i\theta\sum_{i}\sigma_{z}^i)$; $O_3(\theta,\phi)=\exp(-i\theta\sum_{i}\sigma_{\phi}^i)$; $O_4(\theta,\phi)=\exp(-i\theta\sum_{i< j}\sigma^{i}_{\phi}\sigma^{j}_{\phi})$. Here $\sigma_{\phi}{=}\cos{\phi}\sigma_x+\sin{\phi}\sigma_y$ and $\sigma^{j}_{k}$ denotes the $k$-th Pauli matrix acting on the $j$-th qubit. 
$O_4$ is an effective qubit-qubit interaction mediated by a common vibrational mode of the ion string~\cite{Sorensen:1999}. Recent advances have seen the quality of these operations increase appreciably~\cite{Benhelm:2008b}. 
For our  simulations, we define dimensionless Hamiltonians $\tilde{H}$, i.e. $H{=}E\tilde{H}$ such that $U{=}e^{-i\tilde{H}Et/\hbar}$ and the system evolution is quantified by a unitless phase $\theta{=}Et/\hbar$.

We begin by simulating an Ising system of two interacting spin-1/2 particles, which is an elementary building block of larger and more complex spin models, and was one of the first systems to be simulated with trapped ions following an analog approach \cite{Friedenauer:2008qm,PhysRevLett.92.207901}. The Hamiltonian is given by $\tilde{H}_{\mathrm{Ising}}{=}B(\sigma^1_z+\sigma^2_z) + J\sigma^1_x\sigma^2_x$. The first bracketed term represents the interaction of each spin with a uniform magnetic field in the $z$-direction and the second an interaction between the spins in an orthogonal direction. The interactions do not commute, giving rise to non-trivial dynamics and entangled eigenstates. Each spin is mapped directly to an ionic qubit ($\ket{1}{=}\ket{\!\uparrow}$, $\ket{0}{=}\ket{\!\downarrow}$). The dynamics are implemented with a stroboscopic sequence of $O_2$ and $O_4$ gates, representing the magnetic field and spin-spin evolution operators, respectively. We first simulate a time-independent case $J{=}2B$ which couples the initial state $\ket{\!\uparrow\uparrow}$ to a maximally entangled superposition of  $\ket{\!\!\uparrow\uparrow}$ and  $\ket{\!\!\downarrow\downarrow}$ (Fig~1A). The simulated dynamics converge closer to the exact dynamics as the digital resolution is increased. 
The overall simulation quality is quantified using quantum process tomography (QPT)~\cite{Poyatos:1997}, 
yielding a process fidelity of $91(1)$\% at the finest digital resolution used. In~\cite{online} we show how higher-order Trotter decompositions can be used to achieve more accurate digital approximations with fewer operations.

We now consider a time-dependent case where $J$ increases linearly from 0 to
4$B$ during a total evolution $\theta_t$. 
In the limit $\theta_t{\rightarrow}\infty$, spins initially prepared in the paramagnetic
ground state of the magnetic field ($\ket{{\downarrow\downarrow}}$) will evolve 
adiabatically into the anti-ferromagnetic ground state of the final Hamiltonian: an entangled superposition of the $\sum_j\sigma_x^j$ eigenstates $\ket{{\leftarrow\rightarrow}}_x$ and $\ket{{\rightarrow\leftarrow}}_x$. 
As a demonstration, we approximate the continuous dynamics, for $\theta_t{=}\pi/2$, using a stroboscopic sequence of 24 $O_2$ and $O_4$ gates, 
and measure the populations in the $\sigma_x$ basis (Fig~1B). 
The evolution closely follows the exact case and an entangled state is created ($63(6)\%$ tangle~\cite{andrew}). Full quantum state reconstructions are performed after each digital step, yielding fidelities between the ideal digitised and measured state of at least $91(2)\%$, and overlaps with the instantaneous ground state of no less than $91(2)\%$. Note that the observed oscillation in expectation values is a diabatic effect, as excited states become populated.

More complex systems with additional spin-spin interactions in the $y$ (`XY' model) and $z$ (`XYZ' model) directions can be simulated by reprogramming the operation sequence. The dynamics due to an additional spin-spin interaction in the $y$-direction is simulated by adding another $O_4$ operation to each step of the Ising stroboscopic sequence (with $\phi{=}\pi/2$). A third spin-spin interaction in the $z$-direction is realised by adding an $O_4$ gate sandwiched between a pair of $O_3$ operations set to rotate the reference frame of the qubits. In the simulated dynamics of the initial state $\ket{\!\!\rightarrow\leftarrow}_x$ under each model, for a fixed digital resolution of $\theta/n{=}\pi/16$ and up to 12 trotter steps (Fig~2), up to 24, 48 and 84 gates are used for the Ising, XY and XYZ simulations, respectively. This particular initial state is chosen because the ideal evolution is different for each model. The results show close agreement with the exact dynamics and results from QPT after four digital steps yield process fidelities, with the exact unitary evolution, of 88(1)\%, 85(1)\% and 79(1)\%, for the Ising, XY and XYZ respectively. With perfect operations the Trotter error would be less than 1\% in each case. Note that while analog simulations of Ising models have previously been demonstrated in ion traps~\cite{Friedenauer:2008qm,Kim:2010}, XY and XYZ models have not.

The digital approach allows arbitrary interaction distributions between spins to be programmed. For three-spin systems, we realise various interactions that give rise to the dynamical evolutions of the initial state $\ket{\!\uparrow\uparrow\uparrow}$ (Fig~3). Fig~3A shows a system supporting interactions between all spin pairs with equal strength, and between each spin and a transverse field. The initial state couples equally to $\ket{\!\uparrow\downarrow\downarrow}$, $\ket{\!\downarrow\uparrow\downarrow}$ and $\ket{\!\downarrow\downarrow\uparrow}$, while the strength of the field determines the amplitude and frequency of the dynamics. For the case shown ($J=2B$) an equal superposition of the coupled states is periodically created (an entangled W state~\cite{PhysRevA.62.062314}).  
Fig~3B shows how non-symmetric interaction distributions can be programmed, using sequences of $O_4$ and $O_1$ to add spin-selective interactions. The interaction between one spin pair is dominant. Due to this broken symmetry, one coupled state ($\ket{\!\uparrow\downarrow\downarrow}$) is populated faster than the others, yielding more complex dynamics than in the symmetric case. 
Fig~3C demonstrates the ability to simulate $n$-body interactions;  
specifically $\sigma_{z}^1\sigma_{x}^2\sigma_{x}^3$. 
A clear signature is observed: direct coupling between $\sum_j\sigma_y^j$ eigenstates $\ket{\!\rightarrow\rightarrow\rightarrow}_y$ and $\ket{\!\leftarrow\leftarrow\leftarrow}_y$, periodically producing an entangled GHZ state~\cite{PhysRevA.62.062314}. 
Many-body spin interaction of this kind are an important ingredient in the simulation of systems with strict symmetry requirements~\cite{doi:10.1146/annurev-physchem-032210-103512} or spin models exhibiting topological order \cite{RevModPhys.80.1083}''. Measurements in other bases and simulations of nearest-neighbour and many-body interactions with a transverse field using over 100 gates are presented in~\cite{online}. 

Fig~4A shows the observed dynamics of the 4-spin state $\ket{\!\uparrow\uparrow\uparrow\uparrow}$ under a long-range Ising-type interaction.
The rich structure of the dynamics reflects the increased complexity of the underlying Hamiltonian: oscillation frequencies correspond to the energy gaps in the spectrum. This information can be extracted via a Fourier transform of the data (see~\cite{online}).
Specific energy gaps could be targeted by preparing superpositions of eigenstates via an initial quasi-adiabatic digital evolution~\cite{PhysRevLett.97.050504}, or study the complex non-local correlations generated by this model. 
Fig~4B shows the observed dynamics for our largest simulation: a six spin many-body interaction, which directly couples the states $\ket{\!\uparrow\uparrow\uparrow\uparrow\uparrow\uparrow}$ and $\ket{\!\downarrow\downarrow\downarrow\downarrow\downarrow\downarrow}$, periodically producing a maximally entangled GHZ state. 

Direct quantification of simulation quality for more than two qubits is impractical via QPT: for three qubits, expectation values must be measured for 1728 experimental configurations, and this increases exponentially with qubit number ($\approx 3\times10^6$ for six qubits). However, the average process fidelity ($F_p$) can be bounded more efficiently~\cite{Hofmann:2005}. We demonstrate this for the 3- and 6-spin simulations of Fig~3C and 4B, respectively. Bounds of  $85(1)\%{\leq} F_p{\leq} 91(1)\%$ (3 spins) and $56(1)\%{\leq} F_p{\leq} 77(1)\%$  (six spins) are obtained at $\phi{=}0.25$, using 40 and 512 experimental configurations respectively~\cite{online}. The largest system for which a process fidelity has previously been measured is 3 qubits~\cite{PhysRevLett.102.040501}. 
Note that a different measure of process quality is given by the worst-case fidelity, over all input states, and may be better used to assess errors in future full-scale fault-tolerant simulations. Regardless of the measure used, the error in large-scale digital simulations built from finite-sized operations can be efficiently estimated. Each operation can be characterized with a finite number of measurements, then the error in any combination can be chained~\cite{gilchrist:062310}.
In order to exploit this the number of ionic qubits on which our many-qubit operators $O_{2-4}$ can act must be restricted. 

The dominant effect of experimental error can be seen in Fig 3B and 4B: the dynamics damps due to decoherence processes. Laser frequency and ambient magnetic field fluctuations are far from the leading error source: in the absence of coherent operations, qubit lifetimes are over an order of magnitude longer (coherence times ${\approx}~30$~ms) than the duration of experiments (${\approx}~1{-}2$~ms). The current leading sources of error, which limit both the simulation complexity and size, are thought to be laser intensity fluctuations~\cite{online}. This is not currently a fundamental limitation and, once properly addressed, should enable an increase in simulation capabilities.

The digital approach can be combined with existing tools and techniques for analog simulations to expand the range of systems that can be simulated. 
In light of the present work, and current ion trap development~\cite{Home04092009}, digital quantum simulations involving many tens of qubits and hundreds of high-fidelity gates seems feasible in coming years.


%


\noindent\textbf{Acknowledgements}\\
\noindent We thank Wolfgang D\"ur, Al\'an Aspuru-Guzik and Michael Brownnutt for discussions. We acknowledge  financial support by the Austrian Science Fund (FOQUS), the European Commission (AQUTE), the Institut f\"ur Quanteninformation GmbH, IARPA, and two Marie Curie International Incoming Fellowships within the 7th European Community Framework Programme.


\clearpage


\begin{figure}
\begin{center}
\includegraphics[width=9cm]{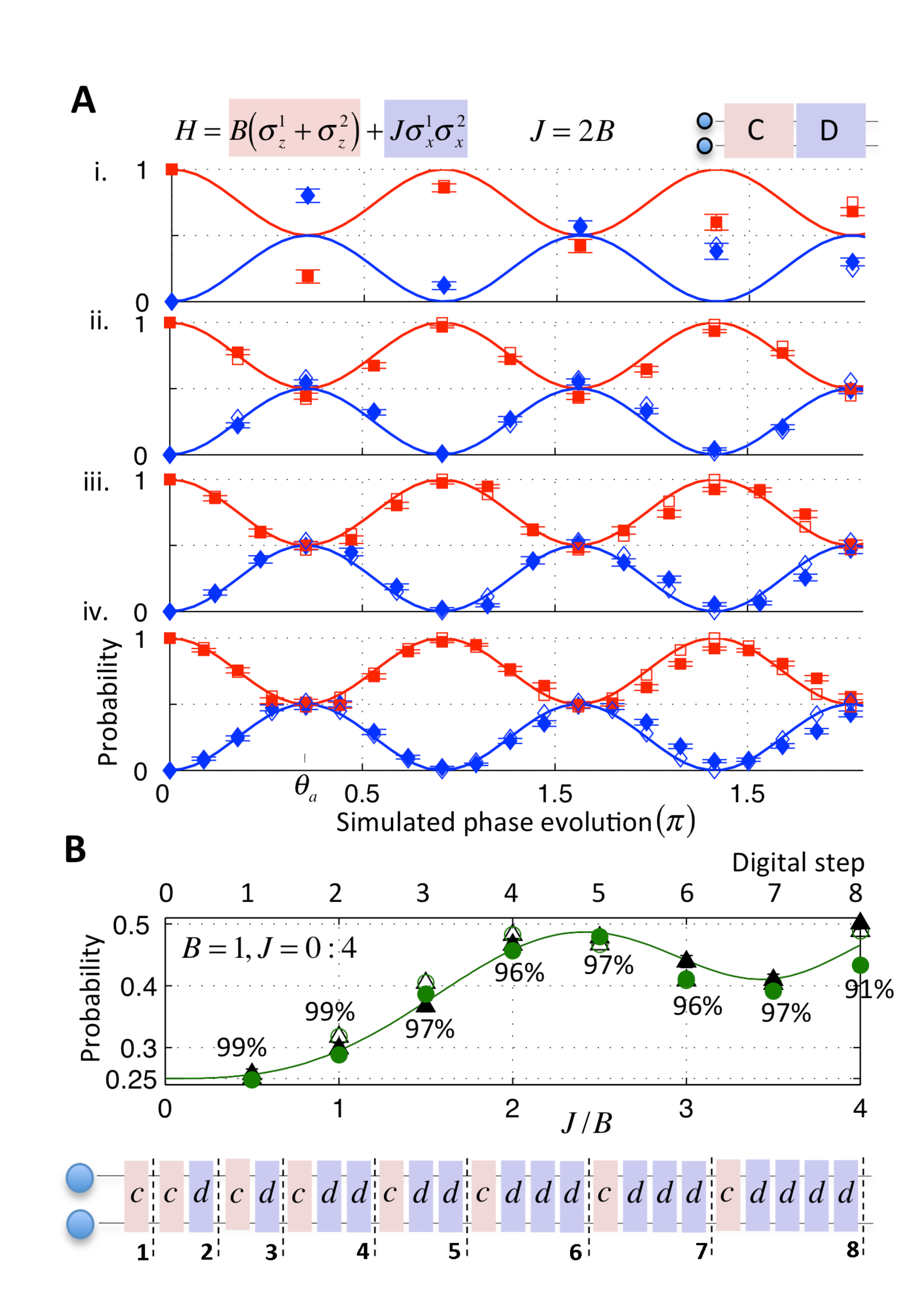}
\end{center}
\caption*{
\noindent {\bf Fig. 1.} Digital simulations of a two-spin Ising system.
Dynamics of initial state $|\!\downarrow\downarrow\rangle$ for two cases.  
\textbf{(A)} a time independent system ($J{=}2B$) and increasing levels of digital resolution (i$\rightarrow$iv.). A single digital step is 
$D.C{=}O_4(\theta_a/n,0).O_2(\theta_a/2n)$,  where $\theta_a{=}\pi/2\sqrt{2}$ and $n{=}1,2,3,4$ (for panels i.-iv respectively). 
Quantum process fidelities between the measured and exact simulation at $\theta_a$ are i) $61(1)\%$ and iv) $91(1)\%$ (ideally 61\% and 98\%, respectively)~\cite{Fey82}.
\textbf{(B)} A time-dependent system. J increases linearly from 0 to 4B. Percentages: fidelities between measured and exact states with uncertainties less than 2\%. The initial and final state have entanglement 0(1)\% and 63(6)\% (tangle~\cite{andrew}), respectively. 
The digitised linear ramp is shown at the bottom: $c{=}O_2(\pi/16)$, $d{=}O_4(\pi/16,0)$. For more details see~\cite{online}.
Lines; exact dynamics. Unfilled shapes; ideal digitised. Filled shapes; data 
($\color{red}{\blacksquare}$$\uparrow\uparrow$ 
$\color{blue}{\Diamondblack}$$\downarrow\downarrow$ 
$\color[rgb]{0, 0.498, 0}{\medbullet}$$\rightarrow\rightarrow_x$ 
$\blacktriangle$$\leftarrow\leftarrow_x$).
}
\label{FIG1}
\end{figure}

\begin{figure}
\begin{center}
\includegraphics[width=0.7 \columnwidth]{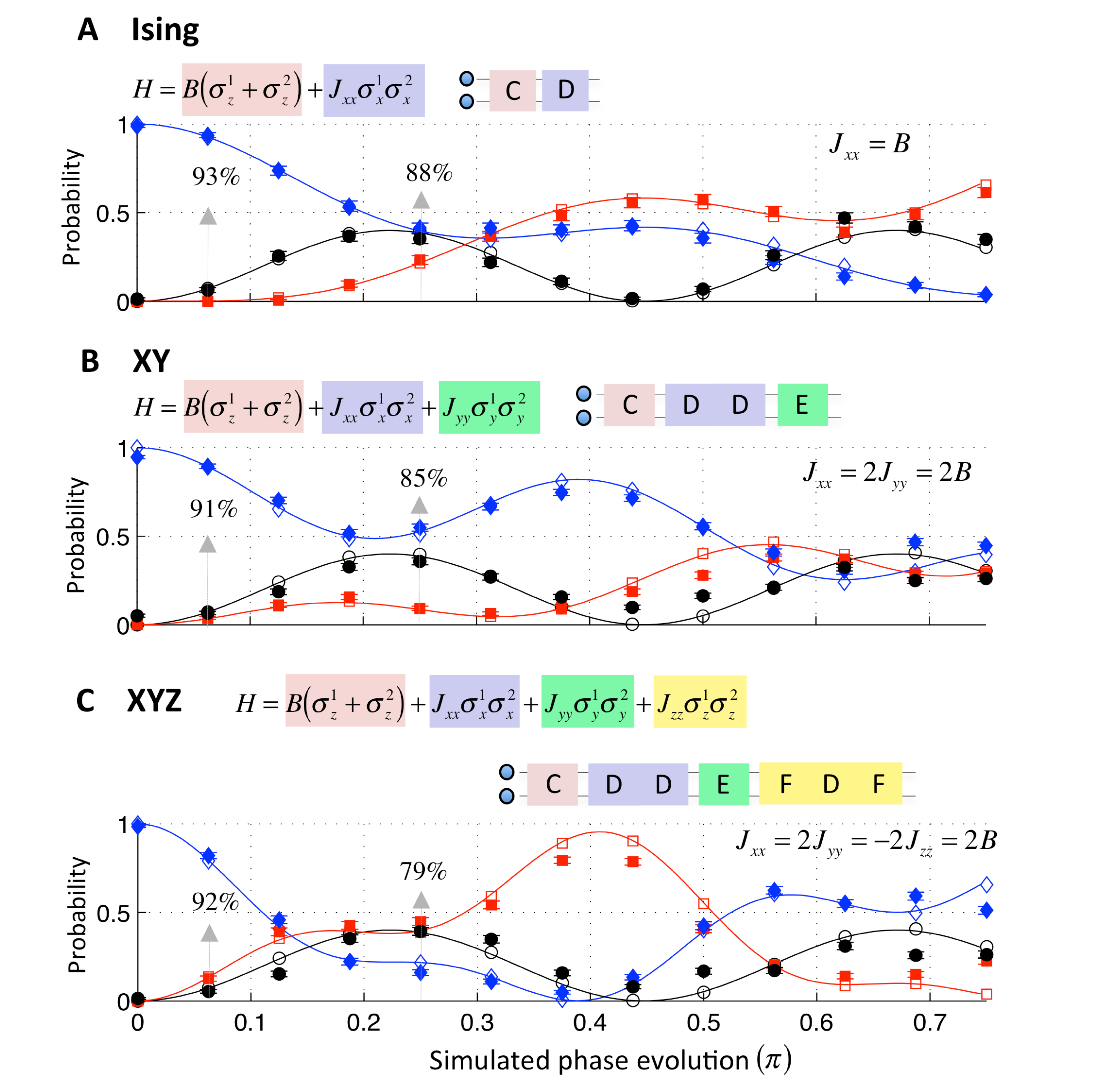}
\end{center}
\caption*{
\noindent {\bf Fig. 2.} Digital simulations of increasingly complex two-spin systems. 
Dynamics of the initial state $\ket{\!\rightarrow\leftarrow}_x$ using a fixed digital resolution of $\pi/16$. The graphic in each panel shows how a single digital step is built:   $C{=}O_2(\pi/16)$, $D{=}O_4(\pi/16,0)$, $E{=}O_4(\pi/16,\pi)$, $F {=}O_3(\pi/4,0)$. 
Quantum process fidelities between the measured and exact simulation after 1 and 4 digital steps are shown with grey arrows (uncertainties $\leq1\%$~\cite{online}). 
Lines; exact dynamics. Unfilled shapes; ideal digitised. Filled shapes; data
($\color{blue}{\Diamondblack}$$\rightarrow\leftarrow_x$ 
$\color{red}{\blacksquare}$$\leftarrow\rightarrow_x$ 
$\blacktriangle$$\leftarrow\leftarrow_x,\rightarrow\rightarrow_x$).
} 
\label{FIG2}
\end{figure}

\begin{figure}
\begin{center}
\includegraphics[width=0.8 \columnwidth]{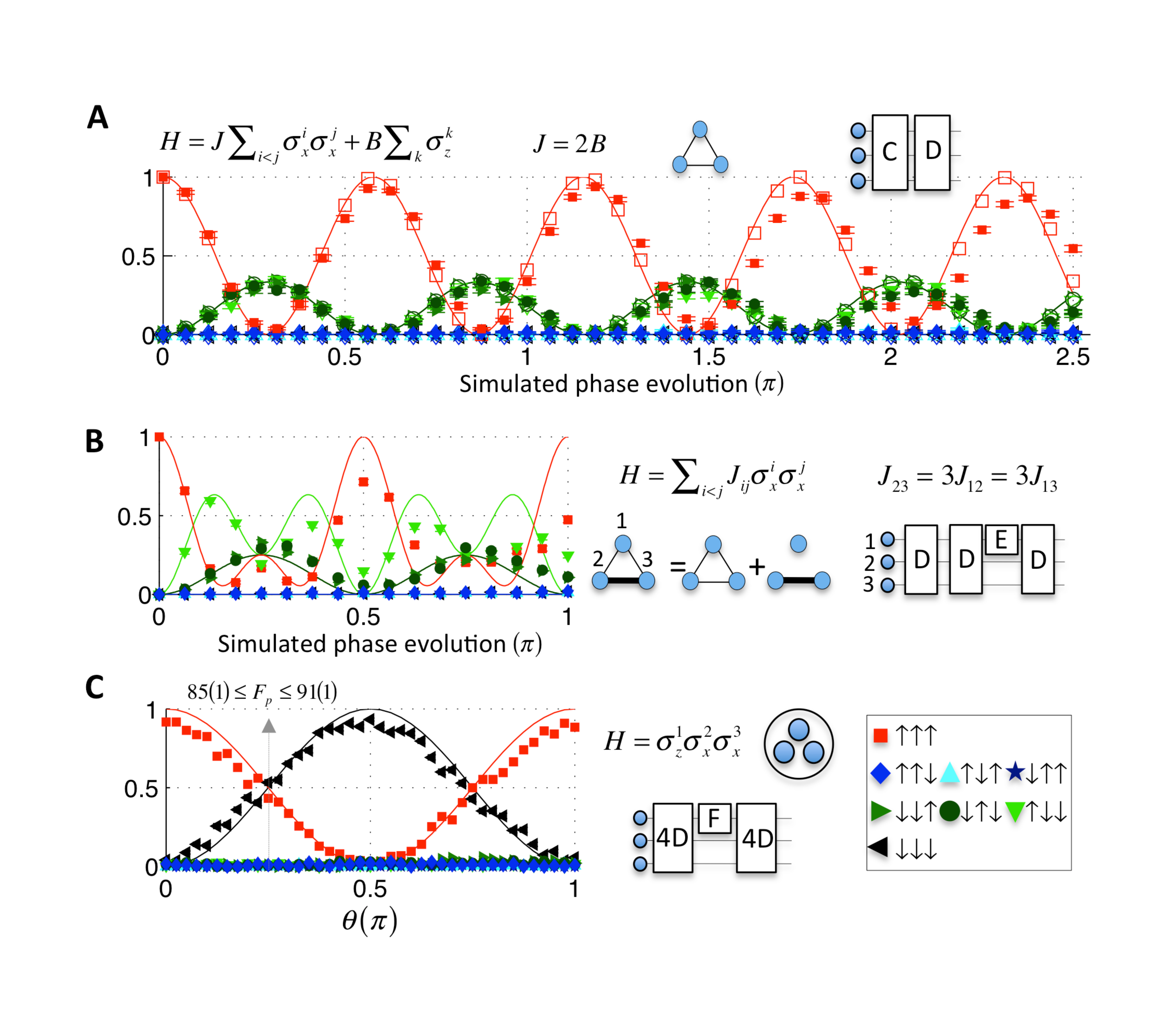}
\end{center}
\caption*{
\noindent {\bf Fig. 3.} Digital simulations of three-spin systems.
Dynamics of the initial state $\ket{\!\uparrow\uparrow\uparrow}$ in three cases. 
\textbf{(A)} Long-range Ising system. Spin-spin coupling between all pairs with equal strength and a transverse field. 
$C{=}O_2(\pi/32)$, $D{=}O_4(\pi/16,0)$. 
\textbf{(B)} Inhomogeneous distribution of spin-spin couplings, decomposed into an equal strength interaction and  another with twice the strength between one pair. $E{=}O_1(\pi/2,1)$. 
\textbf{(C)} Three-body interaction, which couples the $\sum_j\sigma_y^j$ eigenstates $\ket{\!\leftarrow\leftarrow\leftarrow}_y$ and $\ket{\!\rightarrow\rightarrow\rightarrow}_y$. 
An $O_3(\pi/4,0)$ operation before measurement rotates the state into the logical $\sigma_z$ basis. 
$F {=}O_1(\theta,1)$, $4D{=}O_4(\pi/4,0)$. 
Any point in the phase evolution is simulated by varying the phase $\theta$ of operation $F$. Inequalities bound the quantum process fidelity $F_p$  (see~\cite{online} for details).
}
\label{FIG3}
\end{figure}

\begin{figure}
\begin{center}
\includegraphics[width=0.8 \columnwidth]{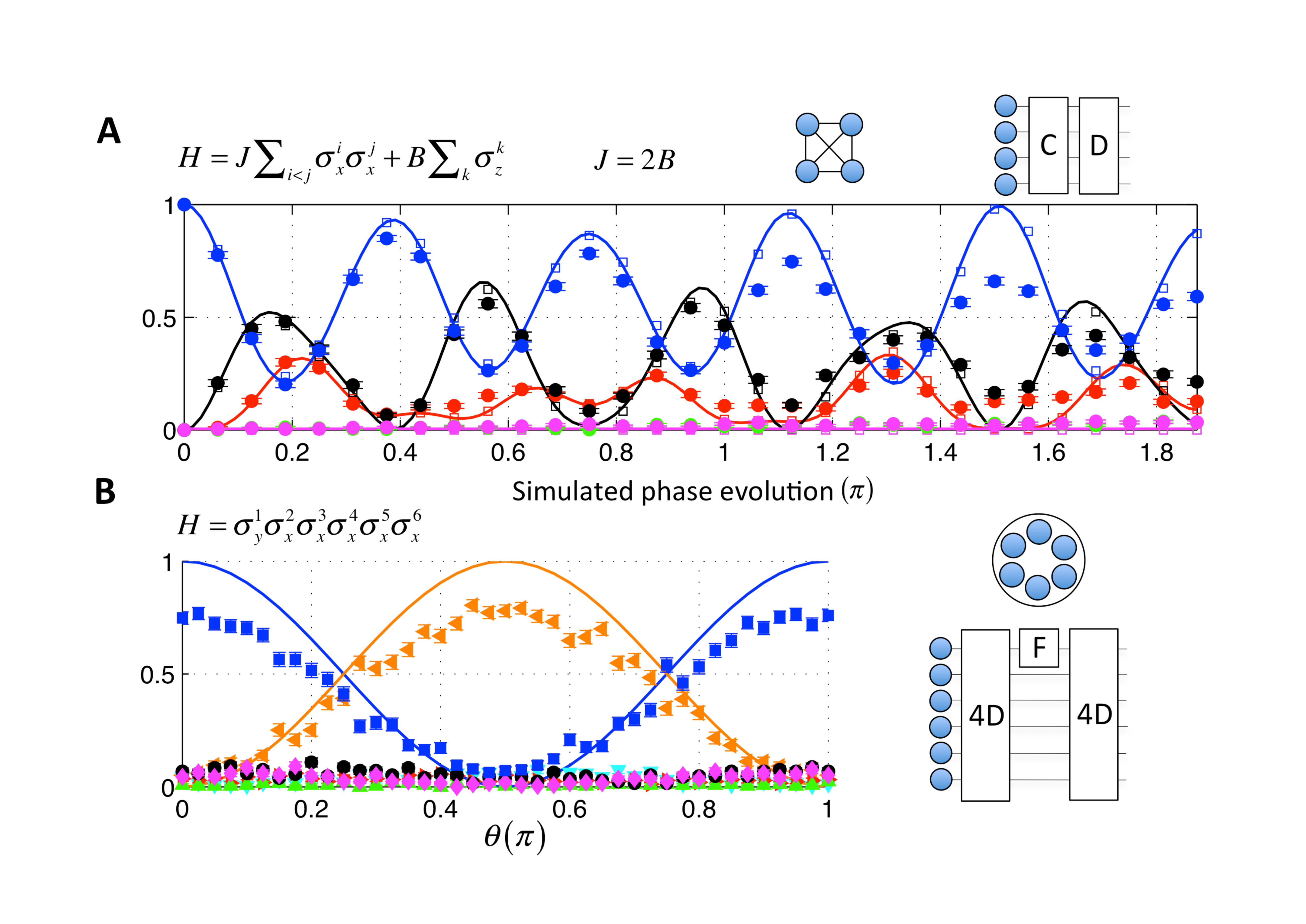}
\end{center}
\caption*{
\noindent {\bf Fig. 4.} 
Digital simulations of four and six spin systems.
Dynamics of the initial state where all spins point up in two cases. 
\textbf{(A)} Four spin long-range Ising system. Each digital step is $D.C{=}O_4(\pi/16,0).O_2(\pi/32)$.
Error bars are smaller than point size.
\textbf{(B)} Six spin six-body interaction.  $F {=}O_1(\theta,1)$, $4D{=}O_4(\pi/4,0)$. 
The inequality at $\phi{=}0.25$ bounds the quantum process fidelity $F_p$ at $\theta{=}0.25$ (see~\cite{online} for details). 
Lines; exact dynamics. Unfilled shapes; ideal digitised. Filled shapes; data 
($\color{blue}{\blacksquare}$$P_0$
$\color{magenta}{\Diamondblack}$$P_1$ 
$\color{black}{\medbullet}$$P_2$
$\color{green}\blacktriangle$$P_3$
$\color{red}\blacktriangleright$$P_4$
$\color{cyan}\blacktriangledown$$P_5$
$\color[rgb]{1, 0.5, 0}\blacktriangleleft$$P_6$, 
where $P_i$ is the total probability of finding $i$ spins pointing down.)
}
\label{FIG4}
\end{figure}

\clearpage

\section*{Supplementary Online Material}

\section{Experimental details}

For each simulation a string of $^{40}$Ca$^+$ ions is loaded into a linear Paul trap. Qubits are encoded in two internal states of each ion. We use the meta-stable (lifetime $\sim$~1~s) $|\!\uparrow\rangle=|D_{5/2},m_j=3/2\rangle$ state and the $|\!\downarrow\rangle=|S_{1/2},m_j=1/2\rangle$ ground state, where $m_j$ is the magnetic quantum number, for the experiments with 2 ions and $|\!\uparrow\rangle=|D_{5/2},m_j=-1/2\rangle$, $|\!\downarrow\rangle=|S_{1/2},m_j=-1/2\rangle$ for the experiments with more ions. These states are connected via an electric quadrupole transition at 729~nm and an ultra-stable laser  is used to perform qubit operations. At the start of each experiment the ions are Doppler cooled on the $S_{1/2}\leftrightarrow P_{1/2}$ transition at 397~nm. Optical pumping and resolved-sideband cooling on the $|\!\downarrow\rangle \leftrightarrow |D_{5/2},m_j=5/2\rangle$ transition prepare the ions in the state $|\!\downarrow\rangle$ and in the ground state of the axial center-of-mass vibrational mode.

The internal states of the ions are measured by collecting fluorescence light on the $S_{1/2}\leftrightarrow P_{1/2}$ transition on a photomultiplier tube and/or CCD camera. Instances where fluorescence light is detected correspond to the ion being in the state $|\!\downarrow\rangle$, instances where it is not correspond to the ion being in state $|\!\uparrow\rangle$. Each simulation experiment (consisting of cooling, state preparation, simulated time evolution and detection) is repeated many times to obtain enough statistics (at least 200 times per data point). The photomultiplier measures the collective fluorescence state of the ion string, from which the probability for any number of ions in the string to be bright can be extracted i.e. $P_0$ (probability for 0 ions being bright), $P_1$ (probability for any 1 ion to be bright) etc. More information can be obtained by using the CCD camera.  By defining regions of interest on the camera sensor, the fluorescence of each ion can be measured individually, from which the logical state of any individual ion can be extracted.  Measurements in other bases are achieved by applying single-qubit operations to the ions before measurement, to map eigenstates of the desired observable to the logical eigenstates.

More detailed information about our experimental setup and techniques can be found in the Ph.D. thesis of Gerhard Kirchmair~\cite{gerhard}.

Quantum simulations are carried out in two different ion traps. Both are based on linear Paul traps of similar design and working with only slightly different experimental parameters. Trap~1 has trapping frequencies $\omega_{\mathrm{ax}}= 2\pi~\times$~1.2~MHz and $\omega_{\mathrm{rad}}= 2\pi~\times~$3~MHz in the axial and radial directions and operates at a magnetic field of 4~G. Trap~2 has trapping frequencies $\omega_{\mathrm{ax}}= 2\pi~\times~$~1.1~MHz and $\omega_{\mathrm{rad}}= 2\pi~\times~$~3~MHz in the axial and radial directions and operates at a magnetic field of 3.4~G.

Trap~2 offers the possibility of trapping larger strings of ions due to an improved vacuum quality and all experiments involving more than two ions were carried out in this trap.

\subsection{Universal operation set}

In this section we explain how the set of operations used in the experiments are implemented. To recap, the operations are

\vspace{-5mm}
\begin{eqnarray}
O_1(\theta,i)&=&\exp(-i\theta\sigma^{i}_{z})\\
O_2(\theta)&=&\exp(-i\theta\sum\limits_{i}\sigma_{z}^i)\\
O_3(\theta,\phi)&=&\exp(-i\theta\sum\limits_{i}\sigma_{\phi}^i)\\
O_4(\theta,\phi)&=&\exp(-i\theta\sum\limits_{i< j}\sigma^{i}_{\phi}\sigma^{j}_{\phi})
\end{eqnarray}
\vspace{-3mm}

\noindent where $\sigma_{\phi}{=}\cos{\phi}~\sigma_x+\sin{\phi}~\sigma_y$ and $\sigma^{j}_{k}$ denotes the $k$-th Pauli matrix acting on the $j$-th qubit.

Each operation is implemented using one of two laser beam paths, which impinge on the ion string from different directions. One beam illuminates all ions equally and can be used to perform global spin rotations on all ions. This is referred to as the `global beam'. The direction of this beam has a large overlap with the axis of the ion string, such that it can couple to the axial motional state. The other beam is tightly focussed, impinges at a 90 degree (68 for trap 2) angle to the ion string and can be used to address ions individually. This is referred to as the `addressed beam'. The particular ion illuminated by the addressed beam can be changed using an electro-optic deflector in $\approx$ 30 $\mu s$.

In our experiments the coupling between the $j^{th}$ ionic qubit and a laser beam at a frequency that is resonant with the qubit transition is well described by the Hamiltonian $H_3{=}\hbar\Omega \sigma_{\phi}^{j}$.
Here, $\Omega$ is the Rabi frequency which represents the coupling strength between the ion and the laser field. $O_3$ is realised via such a resonant laser pulse using the global beam, where $\theta_3{=}\Omega t$, $\phi_3$ is given by the laser phase and $t$ is the duration of the pulse i.e.
$O_3{=}e^{-i\sum_iH_3t}$.

The interaction between an ionic qubit and a beam at a frequency detuned from resonance by $\Delta \gg \Omega$ is given by the Hamiltonian $H_2{=}\hbar\Omega^2/(4\Delta) \sigma_z^{i}$, describing an AC-Stark effect caused by off-resonant coupling to the $S_{1/2}$-$D_{5/2}$ transition and to very off-resonantly driven dipole transitions. $O_2$ is realised by such a detuned laser pulse using the global beam, with $\theta_3{=}\Omega t$ and $t$ is the duration of the pulse i.e.
$O_2{=}e^{-i\sum_i H_2t}$.
This same interaction is used to create $O_1$ by using an addressed beam.

$O_4$ is an effective qubit-qubit (spin-spin) interaction generated via a M{\o}lmer-S{\o}rensen type interaction~\cite{Sorensen:1999}. For this, the global beam is used with a bichromatic laser pulse, whose frequencies are detuned from the carrier by $\pm \omega_{\mathrm{ax}} \mp \delta$, where $\omega_{\mathrm{ax}}$ is the frequency of the axial centre of mass vibrational mode ($\approx$1.2~MHz) and $\delta$ is between 10 and 100~kHz depending on the simulation. After a time $t_{\mathrm{MS}}=1/\delta$, this generates the unitary $O_4$, with $\theta_4=\eta^2\Omega^2/\delta^2$, and $\phi$ given by the sum of the phases of the two light fields~\cite{Roos:2008}. The unitaries in equations 1-4 form a universal set of operations~\cite{Nebendahl:2009}---any arbitrary unitary qubit evolution can be implemented using only these operations.

We note that by choosing $O_1$ to be a far detuned beam, there are no phase stability requirements between the global and addressed laser paths.

\section{Two-spin simulations}

\subsection{Ising demonstration}

\noindent \subsubsection{Time-independent dynamics}

Figure~1A, in the main text, shows results from digital simulations of a time-independent case of a two-spin Ising system, for increasing levels of digital approximation. Each simulation corresponded to a stroboscopic sequence of $O_2$ and $O_4$ operations, where the former simulates an interaction with an external field, and the latter an orthogonal spin-spin interaction. The laser power was set such that $O_2(\pi)$ pulses took $\approx 30\mu$s. The shorter phase evolutions required for each simulation were achieved by varying the pulse length to the correct fraction of this time. For figures~1A i. - iv. these fractions were $\pi/(4\sqrt{2}), \pi/(8\sqrt{2}), \pi/(12\sqrt{2}), \pi/(16\sqrt{2})$, respectively. We note that these fractions relate to a point in the evolution of the simulated system where a maximally entangled state is created ($\pi/2\sqrt{2}$). The pulse length $t_{MS}$ of the $O_4$ operations is set by the laser detuning $\delta$ (see section \textbf{1.1}). For figures~1A i. - iv. detunings were chosen that yield operation times of 120, 60, 40 and 30$~\mu$s respectively. The power of the laser was adjusted to realise the required phase evolution in each case. Varying the operation lengths in this way enabled the total simulation time to be kept constant (up to small changes due to the $O_3$ operation) at $\approx 600~\mu$s.

\begin{figure}
\begin{center}
\includegraphics[width=17cm]{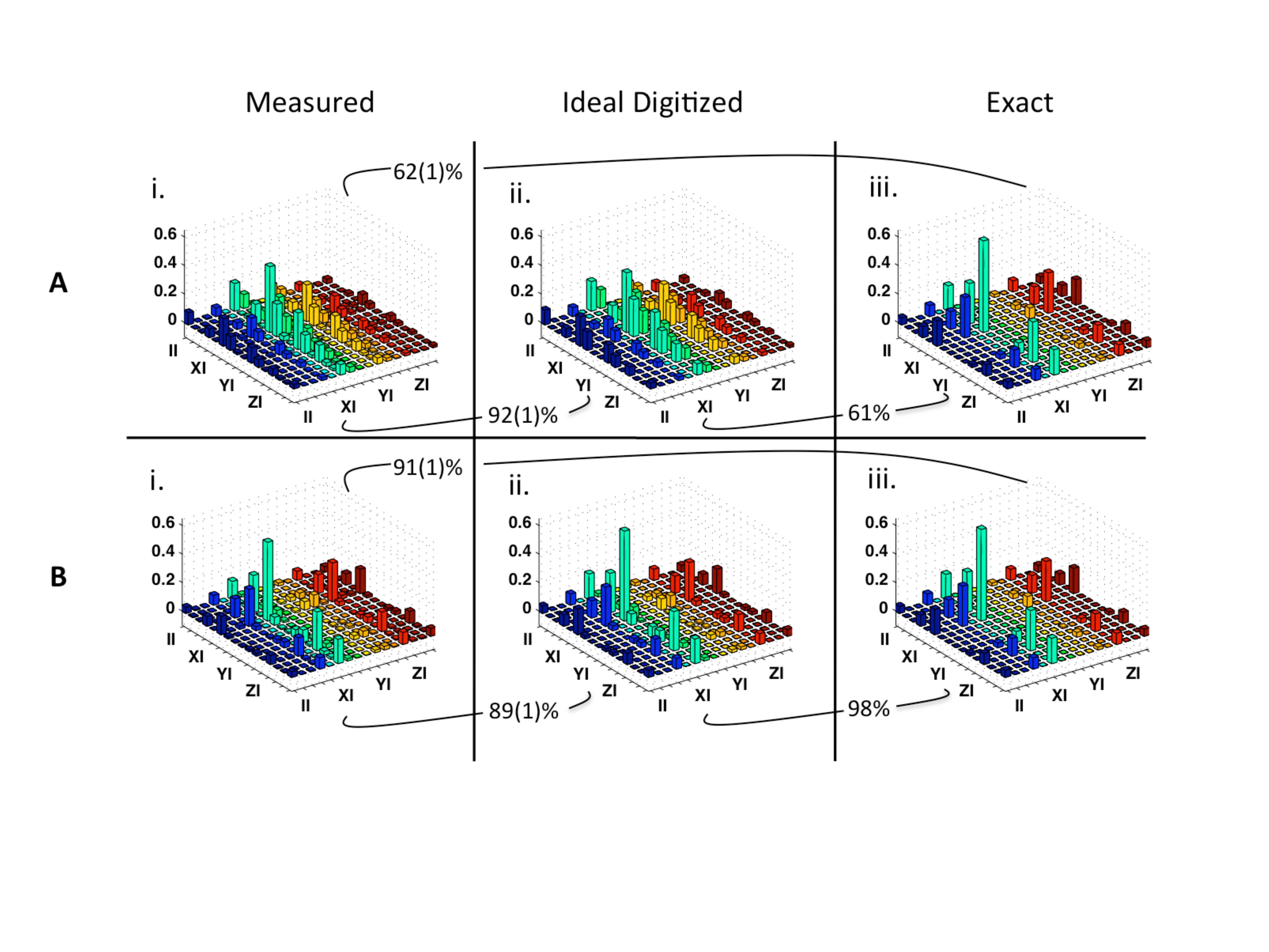}
\end{center}
\vspace{10mm}
\caption{
\textbf{Quantum process matrices of a time-independent Ising-model simulation.} These results support the data shown in Fig.~1A of the main text. The measured processes are digital simulations of the two-spin Ising Hamiltonian $H{=}B(\sigma_z^1+\sigma_z^2){+}J\sigma_x^1\sigma_x^2$ for a phase evolution of $\theta_a{=}\pi/2\sqrt{2}$, and $B=J/2$, using \textbf{(A)} a single Trotter step ($n{=}1$) corresponding to the operation sequence $O_2(\theta_a/2)O_4(\theta_a,0)$ \textbf{(B)} four Trotter steps ($n{=}4$) corresponding to the operation sequence $(O_2(\theta_a/8)O_4(\theta_a/4,0))^4$. Absolute values of experimentally reconstructed process matrices are shown, in the Pauli basis, where $X,Y,Z$ are the usual Pauli matrices and $I$ is the 1 qubit identity operator. Only every fourth operator basis label is shown on the axes to reduce clutter, which go as $II, IX, IY, IZ , XI, XX,XY,....,ZZ$. Percentages are the fidelities between linked matrices, calculated using the full, complex, process matrix and the mixed-state fidelity~\cite{jozsa}. One standard deviation of uncertainty is given in brackets, calculated using Monte Carlo error analysis. The exact matrix is the same in both cases, and is simply the unitary process $exp(-iH\theta_a)$.
}
\label{S1}
\end{figure}

The caption of figure~1A quotes quantum process fidelities for the lowest and highest resolution digital simulations. These are calculated after performing full quantum process tomography, which allows the quantum process matrix to be reconstructed for a given point in the simulation.  For details on process tomography we refer to references ~\cite{Poyatos:1997,Riebe:2006}. In summary we input a complete set of states into the simulation, and for each measure the output in a complete basis. A maximum-likelihood reconstruction algorithm is used to determine the most likely quantum process to have produced our data. Monte-Carlo error analysis is then employed to estimate the uncertainty in the process. Figure \ref{S1} shows the experimentally reconstructed quantum process matrices after a simulated phase evolution of $\theta_\mathrm{a}{=}\pi/2\sqrt{2}$. The matrices clearly demonstrate the improved simulation quality at higher digital resolution, and that our simulations are of high quality across the full Hilbert space.

\subsubsection{Time-dependent dynamics}

Figure~1B in the main text shows results from a digital simulation of a time-dependent two-spin Ising model. The spins are first prepared in the ground state of the external magnetic field, the simulated dynamics corresponds to slowly increasing the strength of the spin-spin interaction such that the state evolves to an approximation of the joint ground state, which is highly entangled. The continuous dynamics are approximated by an 8 step digital simulation built from $O_2(\pi/16)$ and $O_4(\pi/16,0)$ operations, of 10~$\mu$s and 30~$\mu$s duration respectively.

\begin{figure}
\begin{center}
\includegraphics[width=17cm]{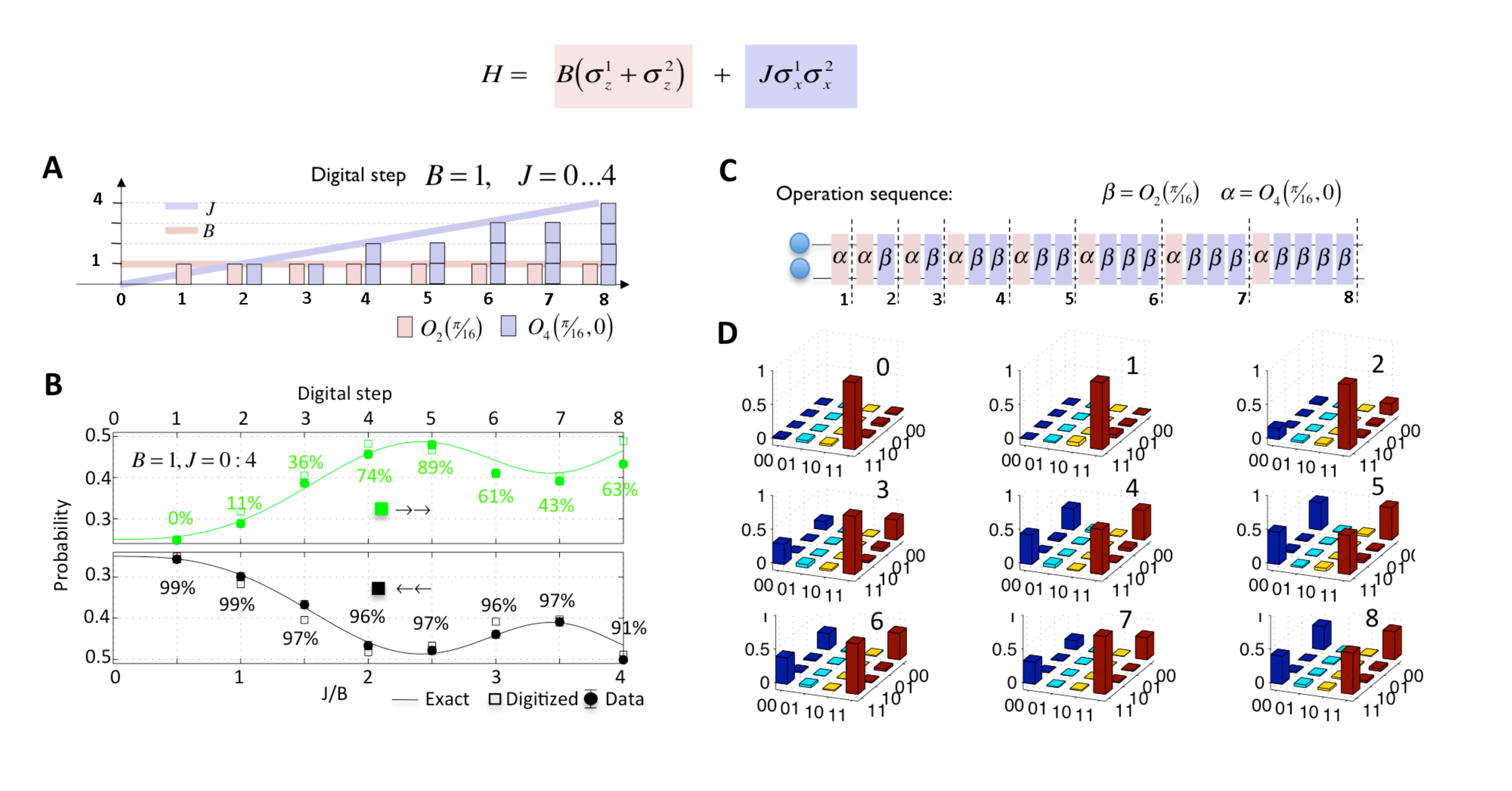}
\end{center}
\caption{\label{timedependantstatetomos}
\textbf{Digital simulation of a time-dependent Ising model.}
These results support the data shown in Fig.~1B of the main text.
\textbf{(A)} A linearly increasing spin-spin interaction strength (over a total phase $\pi/2$) is discretised into 8 digital steps. Each step of the  simulation is built from fixed sized operational building blocks, as shown.
\textbf{(B)} The initial state $\ket{\!\uparrow\uparrow}_x$ evolves into an entangled superposition of the states $\ket{\rightarrow\rightarrow}$ and $\ket{\leftarrow\leftarrow}_x$, which is a close approximation of the ground state of the final Hamiltonian. Black percentages are measured fidelities between the measured and ideal digitised states, quantified by the mixed-state fidelity~\cite{jozsa}. Green percentages quantify the measured entanglement by its tangle~\cite{andrew}. Both are derived from full state reconstructions, shown in (D). Uncertainties in these values of 1 standard deviation, derived from a Monte Carlo simulation based on the experimentally obtained density matrices, are determined for steps 0 to 8 to be
(1, 1, 1, 2, 1, 1, 2, 1, 2)\%
for the overlap with the ideally digitised state and
(1, 1, 2, 4, 5, 3, 5, 4, 6)\%
for the tangle, respectively.
\textbf{(C)} Operation sequence of the digital simulation.
\textbf{(D)} Experimentally reconstructed density matrices at each of the 8 digital steps in the simulation (including the initial state at step 0).
}
\label{S2}
\end{figure}

Figure~\ref{S2} reproduces and extends the data and details in the main text, showing experimentally reconstructed density matrices at all 9 stages of the digital simulation (including the initial state). These matrices are constructed via a full quantum state tomography~\cite{Poyatos:1997,Riebe:2006}. Maximum-likelihood tomography is used to assure a physical state and Monte-Carlo analysis is used to estimate errors in derived quantities. The fidelity and entanglement properties quoted in Figure~1B are calculated from these states. The fidelity is between the measured and ideal digital case, assuming perfect operations. The entanglement is quantified by the tangle which can be readily calculated from the density matrices~\cite{andrew}.

\subsubsection{Higher-order Trotter approximation}

\begin{figure}
\begin{center}
\includegraphics[width=10cm]{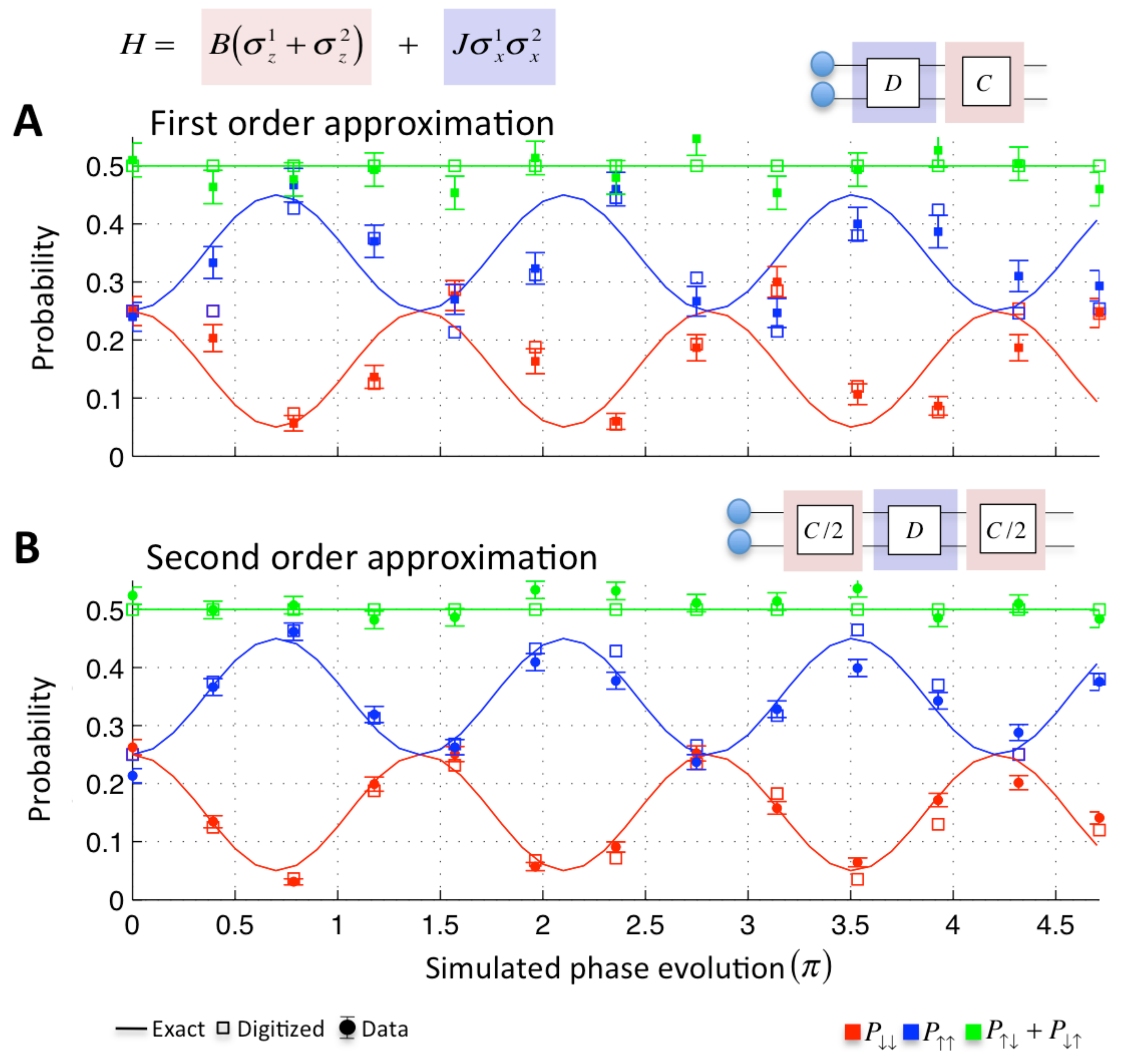}
\end{center}
\caption{
\textbf{First and second-order digital simulations of a two-spin Ising model}.
The graphic in the top right corner of each panel shows the operational sequence for a single digital time step. Dynamics of the initial state $\ket{\!\rightarrow\rightarrow}_x$ using \textbf{(A)} first and \textbf{(B)} second order Trotter-Suziki approximations. $D{=}O_4(\pi/8,0)$, $C{=}O_2(\pi/8)$, $C/2{=}O_2(\pi/16)$, and $B=J$. The initial spin state is created starting from $\ket{\!\downarrow\downarrow}$ and applying an $O_3(\pi/4,\pi/2)$ operation.
}
\vspace{-2mm}
\label{S3}
\end{figure}

Consider a Hamiltonian with two terms $H{=}A{+}B$. A first-order Trotter approximation is:

\begin{equation}
U(t){=}\left( e^{-iAt/n}e^{-iBt/n}\right) ^n+\mathcal{O}(t^2/n)
\end{equation}

\noindent which has errors on the order of $t^2/n$. A second-order Trotter-Suzuki approximation~\cite{Suzuki, NC01} is:

\begin{equation}
U(t)\approx\left(e^{-iBt/2n} e^{-iAt/n}e^{-iBt/2n}\right) ^n+\mathcal{O}(t^3/n)
\label{D}
\end{equation}

\noindent which has errors on the order of $t^3/n$. By splitting the second evolution operator into two pieces and rearranging the sequence, a closer approximation to the correct dynamics is achieved.
Practically this means that a more accurate digital approximation can be achieved at the expense of more operations, but for the same total phase evolution for each step. To illustrate this concept we performed a digital simulation of the two-spin Ising model for $B{=}J{=}1$, using both first- and second-order approximations. Figure~\ref{S3} shows results and details. For the first-order simulation we use building blocks $O_4(\pi/8,0)$ and $O_2(\pi/8)$, which is seen to poorly reproduce the ideal evolution of the initial state $\ket{\!\uparrow\uparrow}$. For the second-order simulation we split the $O_2(\pi/8)$ operator into two pieces (each $O_2(\pi/16)$) and rearranged the sequence according to Eq.~\ref{D}, thereby achieving a much more accurate simulation. Since each evolution operator is simulated directly with our fundamental operations, the higher-order approximation can be employed with little overhead. However, in the more general case where operators must be constructed, such as in our simulation of the XYZ model, there is some finite overhead with simulating any evolution operator regardless of how short it is. In this case there will be a trade-off between increased digital resolution offered by higher-order approximations and additional experimental error introduced through using more operations.

\subsection{Digital simulations of the XY and XYZ models}

\begin{figure}
\begin{center}
\includegraphics[width=15cm]{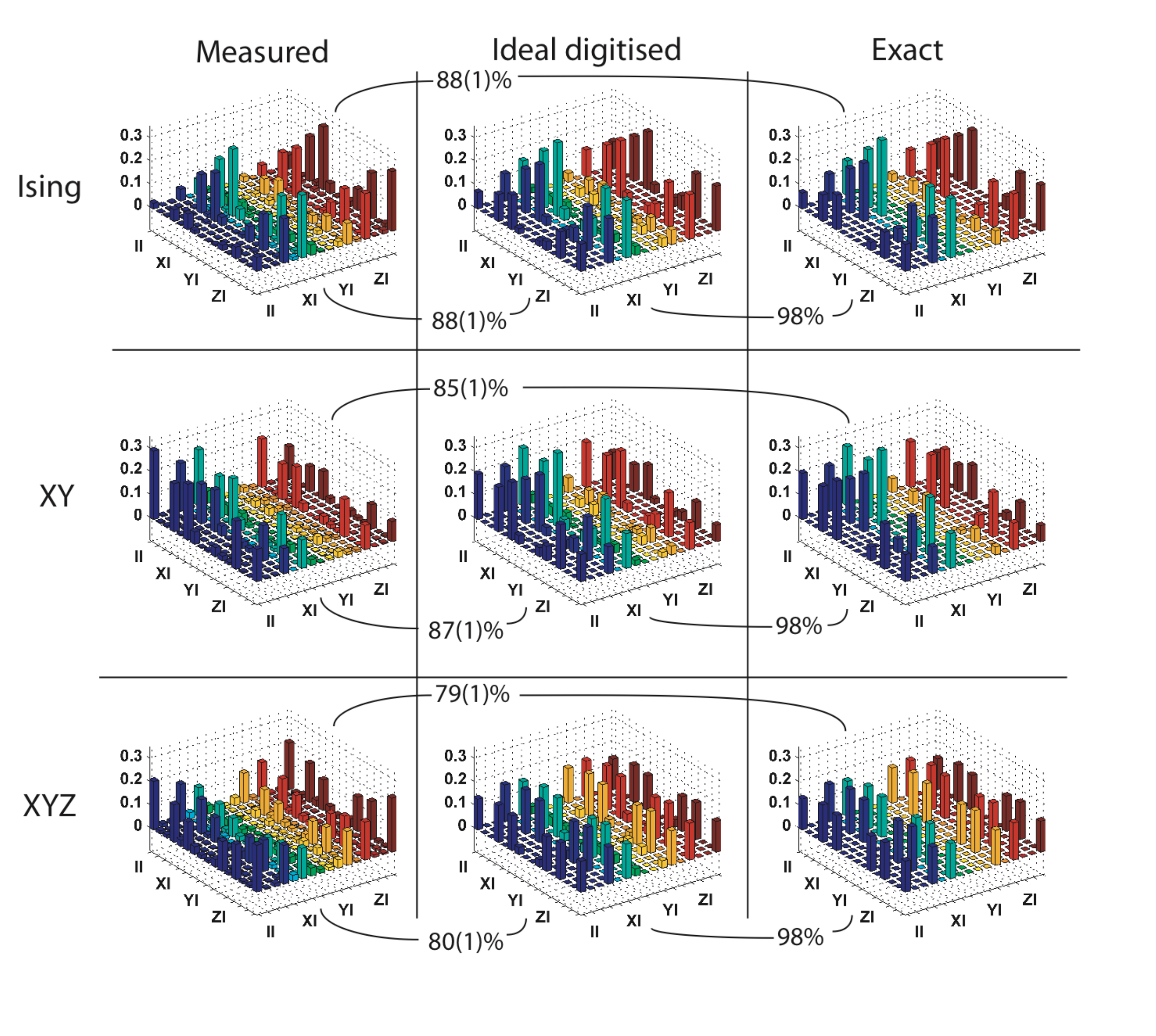}
\end{center}
\vspace{-6mm}
\caption{\textbf{Quantum process matrices of Ising, XY and XYZ simulations.}
Absolute values of experimentally reconstructed process matrices are shown, in the Pauli basis, where $X,Y,Z$ are the usual Pauli matrices and $I$ is the 1 qubit identity operator. Only every fourth operator basis label is shown on the axis to reduce clutter, which go as $II, IX, IY, IZ , XI, XX,XY,....,ZZ$. Percentages are the process fidelities between linked matrices (mixed-state fidelity~\cite{jozsa}). One standard deviation of uncertainty is given in brackets, calculated using Monte Carlo error analysis. The number of fundamental operations implemented for the simulations characterised are 8, 16 and 28 for the Ising, XY and XYZ respectively.
}
\label{S4}
\end{figure}

Figure~2 in the main text shows results of digital simulations of time-independent instances of the Ising, XY and XYZ models. Figure~\ref{S4} in this document shows experimentally reconstructed process matrices of these simulations after 4 digital time steps. This corresponds to a simulated phase evolution of $\theta=\pi/4$. As shown in Figure~2 in the main text, simulations were built from $O_2(\pi/16)$, $O_4(\pi/16,0)$, $O_4(\pi/16,\pi)$, $O_3(\pi/4,0)$ operations, of duration 10, 30, 30, 5~$\mu$s respectively.

\section{Simulations with more than two spins}

\subsection{Ising-type models}

\subsubsection{Long-range}

Our basic set of operations is well suited to simulating the Ising model with long-range interactions:

\begin{equation}
H{=}J\sum_{i<j}\sigma_x^i\sigma_x^j{+}B\sum_k\sigma_z^k
\end{equation}

\noindent which corresponds to a system with interactions between each pair of spins with equal strength J, and a transverse field of strength B. This is because the effective interaction underlying $O_4$ also couples all pairs of spins (ionic qubits) with equal strength. Each digital time step of a simulation requires the sequence $O_4O_2$. In Figure~\ref{S5} we give a more complete set of results for the simulations of three and four spin cases shown in Figures~3A and~4A, of the main text. Specifically, time dynamics measured in a complementary basis and results for different strength transverse fields are shown. The simulated transverse field strength is adjusted by varying the phase evolution of each $O_2$ operation in the digital sequence.

\subsubsection{Aysmmetric and nearest-neighbour}

While our $O_4$ operation is best suited for simulating symmetric interactions between all pairs of qubits (spins), it is possible to engineer interactions that break this symmetry. For this we make use of refocussing techniques in the spirit of nuclear magnetic resonance quantum computing as described in~\cite{Nebendahl:2009}. In particular, we can use the pulse sequence $O_4(\theta/2,\phi)O_1(\pi/2,n)O_4(\theta/2,\phi)$ to exclude ion $n$ from the long-range spin-spin interaction. In principle, sequences of this form can be repeated to simulate any arbitrary spin-spin coupling network.

Two examples with increasing difficulty, in terms of the number of operations required, are given in Figure~\ref{S6}. In Figure~\ref{S6}A the system considered has an asymmetric interaction between the spins: specifically, the interaction strength between one pair is three times larger than any other. A subset of these results is shown in Figure~3B in the main text. Each digital step is constructed from four operations: the first ($O_4$) simulates the evolution due to an interaction between all spin pairs with equal strength for a phase $\theta$, the next three operations simulate the evolution due to an interaction between one pair of spins for a phase $2\theta$. The overall effect is equivalent to evolving the system for a phase $\theta$ due to the desired asymmetric Hamiltonian.

Figure~\ref{S6}B considers a spin system with nearest-neighbour interactions. The large number of operations required for each digital step causes the simulated dynamics to damp due to decoherence processes in the operations themselves (largely the results of laser intensity fluctuations). Clearly these simulations require significantly more gate operations than the long-range Ising model. From an experimental point of view it might therefore be advantageous to use a different set of universal operations for simulating such systems. For this, spin-spin interactions between neighbouring ions could be realised using lasers focussed on pairs of ions. This will be the subject of future work.

\subsection{3-body interaction with additional transverse field}

In the main text we presented simulations of a three-body interaction (Figure~3C). The circuit decomposition for three-body interactions is a special case of a general scheme to simulate $n$-body spin interactions, which is derived and discussed in detail in~\cite{mueller}. We now give simulation results with an additional transverse field. This is particularly challenging as a large number of operations, many of which have large fixed phase evolutions, are required for each digital step. Note that the three-body interaction alone can be simulated for any phase evolution using only three operations, as shown in Figure~3C in the main text. However, the Trotter approximation must be employed in the case of an additional transverse field, costing 4 operations (3 for the three-body interaction and one for the magnetic field interaction) for each digital step of the total phase evolution.  Figure~\ref{S7} shows results, first for the case where the field is zero but following a stroboscopic approach with fixed operation settings, and second with a non-zero field. A coarse digital resolution of $\pi/4$ is chosen so as to observe some dynamics before decoherence mechanisms equally distribute population among each possible spin state.

Note that here, for the first time, we simulate a transverse field using $O_3$ instead of $O_2$. This is because the three-body interaction that we simulate is $\sigma_z^1\sigma_x^2\sigma_x^3$,  and the correct transverse field axis is therefore $\sum_{i}\sigma_y^i$. An alternative approach would be to use two extra pulses on spin 1 to rotate the axis of its spin-spin interaction to the $x$ basis, at the expense of two more operations for each digital step. 

\subsection{Process bounding method}

Quantum process tomography enables a complete reconstruction of the experimental quantum process matrix~\cite{Poyatos:1997, NC01} from which any desired property, such as the process fidelity, can be calculated. However, the number of measurements required grows exponentially with the qubit number. In an ion trap system $12^n$ expectation values must be estimated to reconstruct the process matrix of an $n$ qubit process. This number is already impractical for processes involving more than two qubits: it simply takes too long to carry out the measurements with sufficient precision, while maintaining accurate control over experimental parameters.

In \cite{Hofmann:2005} it is shown that the overall process fidelity can be bounded without reconstructing the process matrix, and with a greatly reduced number of measurements. In summary, the technique requires classical truth tables to be measured for two complementary sets of input basis states. The two sets, $\{\psi_i\}$ and $\{\phi_i\}$, are complementary if $|\langle\psi_i|\phi_i\rangle|^2=1/N$ for all $i$.
Conceptually this means that a measurement in one basis provides no information about the outcome of a subsequent measurement in the other basis. In this way there is no redundancy in these measurements and maximal information is returned.

For a unitary quantum process $U$ a truth table shows the probability for measuring the ideal output state ($\{U\psi_i\}$ and $\{U\phi_i\}$) for each input state in a basis set. 
If we define the fidelity (overlap) of truth-table $i$ with its ideal case as $F_i$, then the process fidelity $F_{p}$ is bound above and below in the following way:

\begin{equation}\label{eq_bound}
F_1+F_2-1\leq F_{p} \leq Min\{F_1,F_2\}
\end{equation}

It is useful to note that the truth table fidelity is equivalent to the average output state fidelity. Therefore, for an $n$ qubit process, the requirements are to prepare two complementary sets of $2^n$ input states, and to measure the probability of obtaining the correct output state (the state fidelity) in each case. The technique is highly dependent on the particular process to be characterised: the challenge is to choose complementary sets of states that can be accurately prepared and for which the output state fidelities can be accurately measured.

\subsubsection{Process bounding the 3-body interaction}

We bounded the process fidelity of the 3-body operation $U_3(\theta)=e^{-i\theta \sigma_z^1\sigma_x^2\sigma_x^3}$ for $\theta{=}\pi/4$ and $\pi/8$, considered in Figure~3C of the main text. As the first basis set we chose the 8 separable eigenstates, i.e. $|0\rangle|++\rangle_x, |0\rangle|+-\rangle_x,..., |1\rangle|--\rangle_x$ (where we now use the conventional qubit state notation for simplicity, and $\ket{\pm}_x=(\ket{0}\pm\ket{1})/\sqrt{2}$). These states can be created experimentally  using a sequence of coherent laser pulses that include both global and addressed beams. The inverse of this pulse sequence, followed by fluorescence detection, is used to effectively perform a projective measurement in this eigenstate basis. From this measurement the average output state fidelity can be calculated directly. The results of these measurements, for $\theta=\pi/4$, are presented in Table~\ref{T1}.

For the second basis set we chose the 8 states $|\!+++\rangle_y, |\!++-\rangle_y,..., |\!---\rangle_y$ (where $\ket{\pm}_y=(\ket{0}\pm i \ket{1})/\sqrt{2}$). 
These input states are mapped to entangled output states by $U_3$.
We then map each output state to a GHZ-like state with the ideal form $\psi_{\mathrm{ideal}}{=}\cos{\theta} |000\rangle+\sin{\theta} |111\rangle$, using an additional set of local operations.
The fidelity of an experimentally produced state $\rho$ (density matrix) with $\psi_{\mathrm{ideal}}$ can be derived from 5 expectation values. The first two are the probability for finding the qubits all in $0$ ($P_{000}$) and the probability for finding them all in $1$ ($P_{111}$).  These can be estimated directly from fluorescence measurements. The last three are the parity of the state (spin correlations) measured in different bases. For this purpose we apply an operation $O_3(\pi/4,\phi)$ to the output state, for three different values of $\phi_{1,2,3}=0, \pi/3, 2\pi/3$ and then measure the parity $\langle\sigma_z^1\sigma_z^2\sigma_z^3\rangle$ via fluorescence measurement and post-processing. It is straightforward to show that the fidelity of an experimentally produced state with the GHZ-like state $\psi_{\mathrm{ideal}}$ is given by

\begin{equation}
	\begin{split}
		&F(\rho,\psi_{\mathrm{ideal}})=\\ & \cos^2 (\theta) P_{000}+\sin^2 (\theta) P_{111}+\frac{\cos \theta \sin \theta}{3}\sum_{i=1}^3\alpha_i q(\phi_i)
	\end{split}
\end{equation}

\noindent where $q(\phi_i)$ is the measured value of the parity at $\phi_i$ and $\alpha_i=\pm 1$ depending on whether the ideally expected parity is at a minimum or at a maximum.

The measurement results for this second set of input states are presented in Table~\ref{T2}.
Together with the results of Table~\ref{T1} and Equation~\ref{eq_bound} the process fidelity can be bound to $0.850(8)\leq F_{process} \leq 0.908(6)$.
This procedure was repeated for $\theta{=}\pi/8$, yielding very similar results: $0.839(9)\leq F_{p} \leq 0.909(7)$.

\subsubsection{Process bounding the 6-body interaction}

We bounded the process fidelity of the 6-body operation $U_6(\theta)=e^{-i\theta \sigma_y^1\sigma_x^2\sigma_x^3\sigma_x^4\sigma_x^5\sigma_6^3}$ for $\theta{=}\pi/4$, shown in figure~4B of the main text. 
Our method is conceptually equivalent to that for the 3-body case described above. As the first basis set of input states we chose the 64 separable eigenstates and directly measured in this basis to extract the output state fidelities. These results are split between Table~\ref{T3} and Table~\ref{T4}. For the second set we chose a complementary basis which evolve into entangled states that are locally equivalent to GHZ states. In the 6-qubit case 2 populations and 6 parities are required for the state fidelity. These results are split between Tables \ref{T5} and \ref{T6}. The fidelity of an experimentally produced state with a GHZ-like state of $n$ qubits ($\Psi{=}\cos{\theta}|0\rangle^{\otimes n}+\sin{\theta}|1\rangle^{\otimes n}$) is given by

\begin{equation}
	\begin{split}
		&F(\rho,\Psi)=\\ & \cos^2 (\theta) P_{00...0}+\sin^2 (\theta) P_{111...1}+\frac{\cos \theta \sin \theta}{n}\sum_{i=1}^n\alpha_i q(\phi_i)
	\end{split}
\end{equation}

where the $n$ values of $\phi$ are equally spaced by $\pi/n$ and alternately correspond to parity maxima ($\alpha=+1$) and minima ($\alpha=-1$). The requirement to measure the parity at $n$ different angles for an $n$-qubit state reflects the increasing number of possible entanglement partitions with $n$.

In total therefore $2^n(n+1)+2^n$ expectation values have to be measured to bound these many-body processes: $2^n(n+1)$ for the basis that becomes entangled and; $2^n$ for the separable eigenbasis. This compares well with the $12^n$ required for full quantum process tomography. For three qubits this means 40 instead of 1728 and for six qubits 512 instead of $2, 985,984$.

Interestingly, further analysis of the data in Tables \ref{T5} and \ref{T6} shows that decoherence of the GHZ states is an error source. The 64 states can be separated into 4 groups, determined by the magnitude of the difference in the number of 0's and 1's in the input state. The possible values are 6, 4, 2, 0. Input states with a difference of 0 (e.g. $\ket{000111}$) are converted, by $U_6$, to states that are eigenstates of $\sigma_z$ rotations on any or all qubits. This makes them free of decoherence effects due to fluctuating magnetic fields, for example. Input states with the maximum difference ($\ket{000000}$ and $\ket{111111}$) are converted to states that are maximally sensitive to these kinds of rotations and errors - by a factor of 6 times more than a single qubit~\cite{monz:2010}. The effect is to reduce the coherence between the populations, which would reduce the parity amplitude while keeping the population values the same. We should expect the average parity amplitude (over the 6 measurements) for the groups 6, 4, 2 and 0 to be progressively better in that order. The results are consistent: the average absolute parity amplitudes for groups 6, 4, 2, 0 are 0.58(2), 0.67(1), 0.71(1) and 0.76(1), respectively, while the average total populations are 0.78(5), 0.83(2), 0.81(1) and 0.83(2), respectively. 

\subsection{Fourier transform to extract energy gaps}

\begin{figure}
\vspace{-30mm}
\begin{center}
\includegraphics[width=13cm]{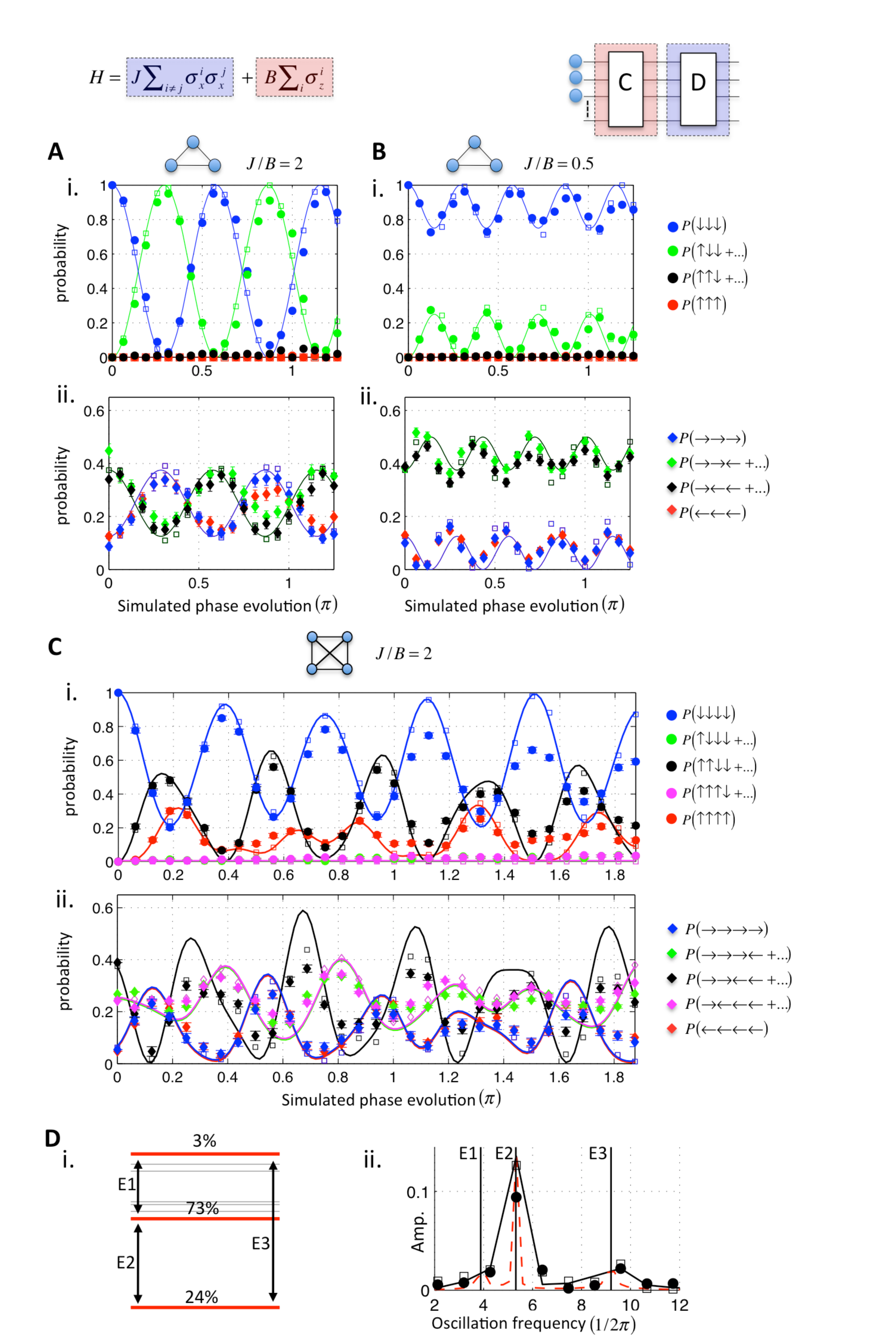}
\end{center}
\vspace{-5mm}
\caption{\textbf{Digital simulations of long-range Ising model.}
A single digital time step in each case is $C.D{=}O_2(B\pi/16,\pi).D{=}O_4(\pi/16,0)$. (--- Exact, $\square$ Ideal Digital, $\medbullet$/$\Diamondblack$ Data).
 \textbf{(A)} 3 spin case for $J/B{=}2$ (A) i. is the same as Fig.3A in the main text.
\textbf{(B)} 3 spin case for $J/B{=}0.5$ such that the transverse field is dominant.
\textbf{(C)} 4 spin case for $J/B{=}2$. (C) i. is the same as Fig.4A in the main text.
\textbf{(D)}. i. Energy level diagram of the 4-spin Hamiltonian simulated in panel C. The initial state $\ket{\!\downarrow\downarrow\downarrow\downarrow}$ is a superposition of 3 of the different energy eigenstates, highlighted in red and with probabilities given as percentages (unpopulated energy levels are shown in grey).
ii. Fourier transform of the black data trace in panel (C) i. clearly resolving the energy gap E2. The dashed red trace is the Fourier transform of the exact dynamics after 4 times the evolution window.
}
\label{S5}
\end{figure}

\begin{figure}
\begin{center}
\includegraphics[width=13cm]{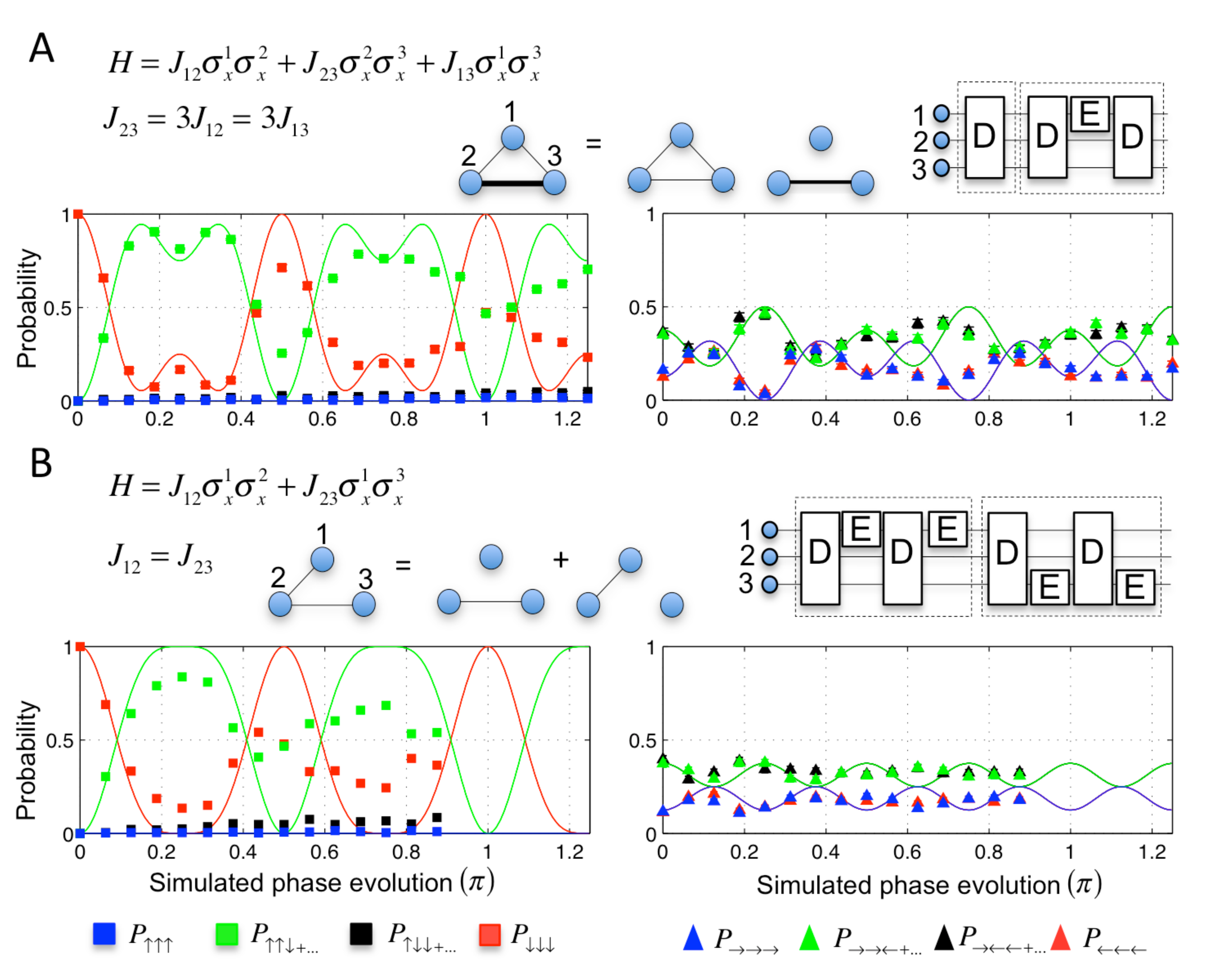}
\end{center}
\vspace{0mm}
\caption{
\textbf{Engineering arbitrary spin-spin coupling distributions}.
Three spin simulations of Ising models with unequal spin-spin coupling distributions.  Results are presented in two complementary bases (--- Exact, $\blacksquare$/$\blacktriangle$ Data).
The graphic in the top right corner of each panel shows the operational sequence for a single digital step. $D{=}O_4(\pi/16,\pi)$, $E{=}O_1(\pi/2,n)$, where $n$ is the qubit on which $E$ is shown to operate in the graphic.
\textbf{(A)} The coupling strength is twice as large between spins 2 and 3 as any other pair. (A)i is the same as Fig.~3A in the main text.
\textbf{(B)} Nearest-neighbour coupling. The first operation sequence $D.E.D.E$ couples only spins 2 and 3. The second sequence couples only spins 1 and 2. We note that the total simulated phase evolution in (B) is less than in (A) simply because the number of pulses that can be stored in the memory of our pulse generator had reached its default limit (150 phase coherent pulses). This could be increased with a small amount of work, but the dynamics can already be seen to have significantly decayed by this point.
While all terms commute in each Hamiltonian, a stroboscopic approach is used, as will be necessary in future simulations where additional non-commuting interactions are incorporated. 
}
\label{S6}
\end{figure}

\begin{figure}
\begin{center}
\includegraphics[width=13cm]{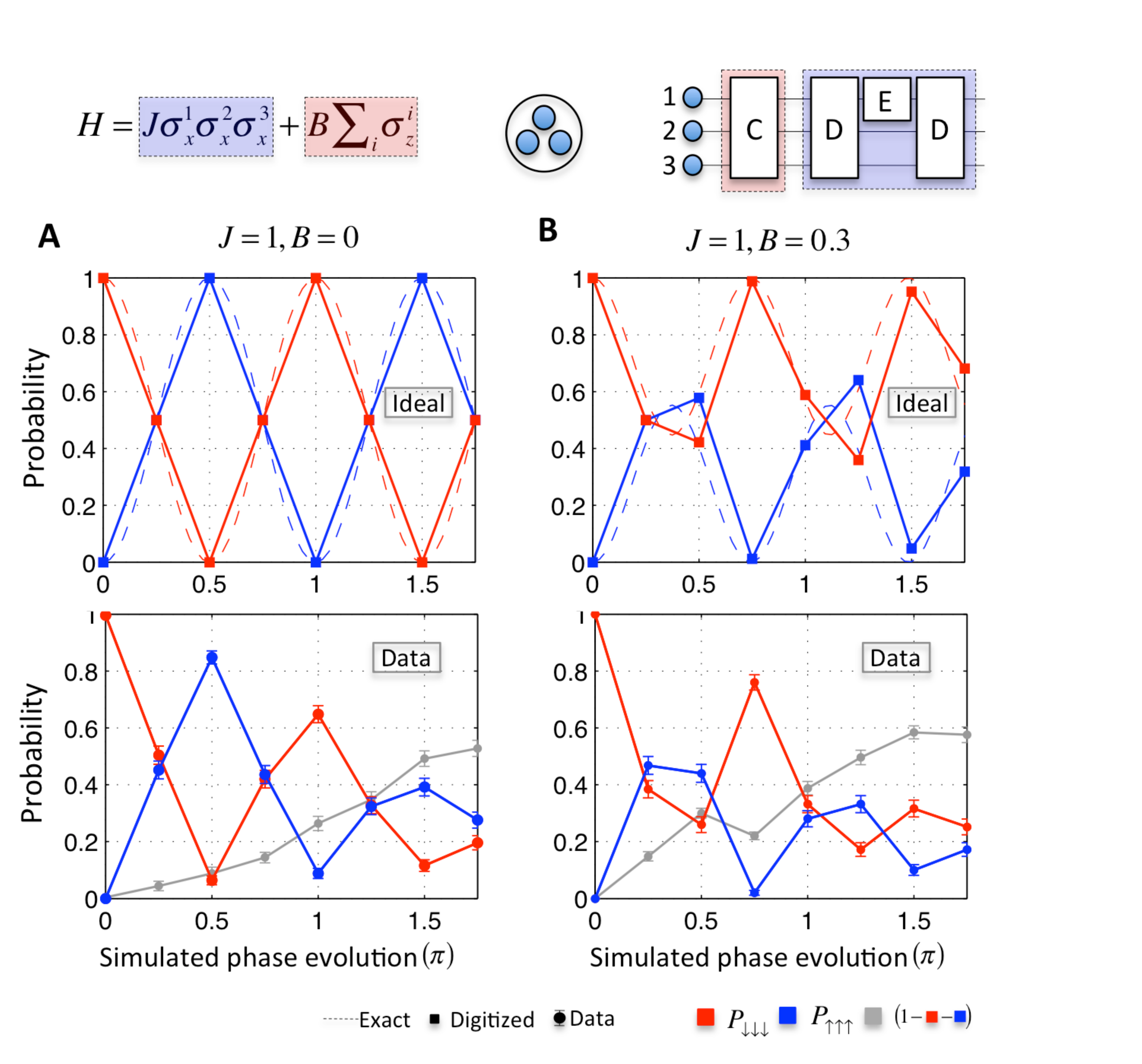}
\end{center}
\vspace{0mm}
\caption{
\textbf{Digital simulations of a three-spin interaction with a transverse field.}
The upper graphs compare exact and ideal-digitized dynamics, the lower graphs show measured results. \textbf{(A)} Zero transverse field.
This reproduces the dynamics shown in Fig.~3C of the main text, but in a stroboscopic way i.e. rather than adjusting the phase of operation $E$ to simulate the dynamics, the phase of $E$ is fixed at $\pi/4$ and the sequence $D.E.D$ is repeated for each data point. This is expensive in terms of the number of operations required, but essential if additional non-commuting interactions are to be simulated following a digital approach.
 \textbf{(B)} Non-zero transverse field.
In both cases the dynamics rapidly damp due to the large number of operations for each digital step and subsequent decoherence.
The graphic in the top right corner shows the operational sequence for a single digital time step. $C{=}O_3(B\pi/4,\pi)$, $D{=}O_4(\pi/4,0)$, $E{=}O_1(\pi/4,1)$.}
\label{S7}
\end{figure}

Oscillation frequencies in the time evolution of observables are energy gaps in underlying Hamiltonian. A Fourier transform can extract this information. We now give a brief example for one of the observables measured in the 4-ion long-range Ising simulation (Figure~\ref{S5}C i. and Figure~4A), which shows the richest dynamics of all our simulations.

Figure~\ref{S5}D shows the spectrum of the ideal Hamiltonian and the probability distribution of the initial state used for the simulation, amongst the energy levels. Three of the nine energy levels are populated, therefore at most 3 energy gaps (oscillation frequencies) can be observed in the dynamics.  However, the observed spectral amplitudes in the Fourier transform depend not only on the population distribution of the initial state, but also the observable coupling strengths and trace length (total simulated phase evolution). Figure~\ref{S5}D shows the Fourier transform of the black data trace in Figure~4A of the main text (and Figure ~\ref{S5}C), which represents  the total probably of finding all combinations of two spins up and two down. One of the three fundamental frequencies is clearly resolved.

\subsection{Error sources in gate operations}

\subsubsection{Laser-ion coupling strength fluctuations}

For previous work on error sources in our quantum operations we refer the reader to~\cite{Kirchmair:2009, Benhelm:2008b, Roos:2008}.  A conclusion from this work is that fluctuations in the laser-ion coupling strength $\Omega$ are a significant source of experimental error. These fluctuations can be introduced by laser-intensity noise or thermally occupied vibrational modes, for example. Since the phase angles $\theta$ of the $O_4$,  $O_1$ and $O_2$ operations are all proportional to $\Omega^2$ this error source is particularly important in our simulations.

We measured the laser intensity fluctuation, at the entrance into the ion-trap vacuum vessel, using a fast photo-diode. A slow oscillation in the intensity by between 1 and 2\% was observed over periods of several minutes. The precise value in the range varies on a daily basis. This corresponds to a coupling-strength fluctuation (intensity is proportional to $\Omega^2$) of approximately 1\%. This measurement is an underestimate of intensity fluctuations at the point of the ions, due to possible beam-pointing fluctuations and wave-front aberrations.

The effect of coupling strength fluctuations on our digitised simulations was modelled, and compared to one of our key results: the time dynamics of a two-spin Ising system at the highest digital resolution used (shown in Figure 1A, panel iv, in the main text). The model makes the assumption that $\Omega$ is constant for each simulation sequence ($\approx 1~ms$), as supported by our photo-diode measurements, but varies from sequence to sequence  (in experiments expectation values are calculated by averaging a large number of repeated experimental sequences taken minutes apart). Fluctuations are incorporated by post-mixing a large number of simulated sequences, with each subjected to noise randomly sampled from a gaussian distribution with a standard deviation $\delta \Omega/\Omega$.

\begin{figure}
\begin{center}
\includegraphics[width=13cm]{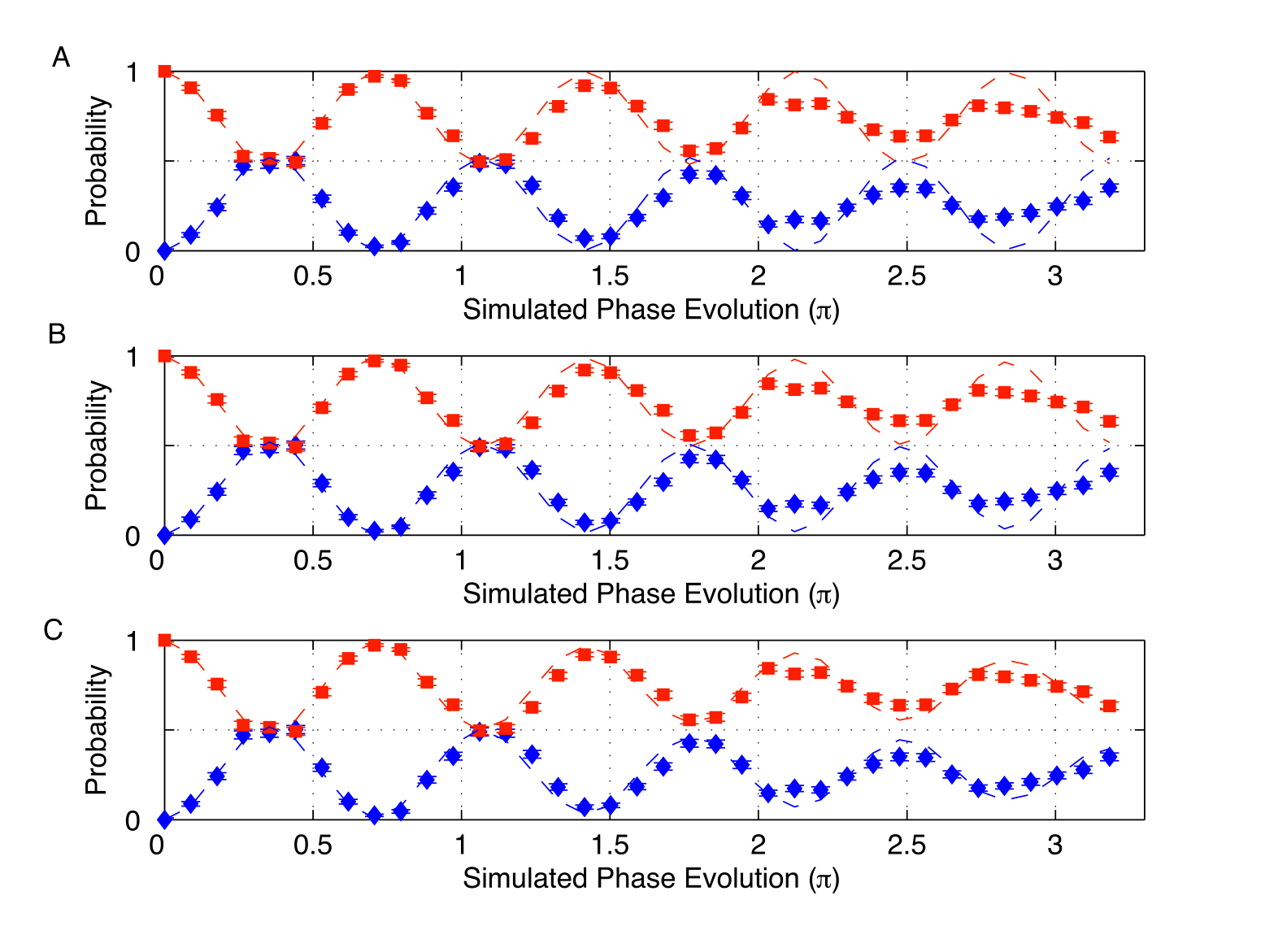}
\end{center}
\vspace{0mm}
\caption{
\textbf{Effect of laser-ion coupling strength fluctuations.}
In each panel simulation results from a two-spin Ising model are compared with a theoretical model that incorporates increasing amounts of fluctuations in laser-ion coupling strength $\Omega$. \textbf{(A)} $d\Omega/\Omega{=}0\%$ \textbf{(B)} $1\%$ \textbf{(C)} $2\%$. 
Dashed lines; predicted results from model. Filled shapes; data. 
($\color{red}{\blacksquare}$$\uparrow\uparrow$ 
$\color{blue}{\Diamondblack}$$\downarrow\downarrow$) 
}
\label{S8}
\end{figure}

Figure~\ref{S8} shows the results: the measured coupling strength fluctuation of 1\% qualitatively reproduces the observed damping of the spin dynamics, while a much closer fit is obtained for a larger fluctuation of 2\%. There are a large number of other errors sources that could contribute to deviations between the observed simulations and the ideal, as discussed in~\cite{Benhelm:2008b, Roos:2008}.

\subsubsection{Frequency shifts in simulated dynamics}

\begin{figure}
\begin{center}
\includegraphics[width=13cm]{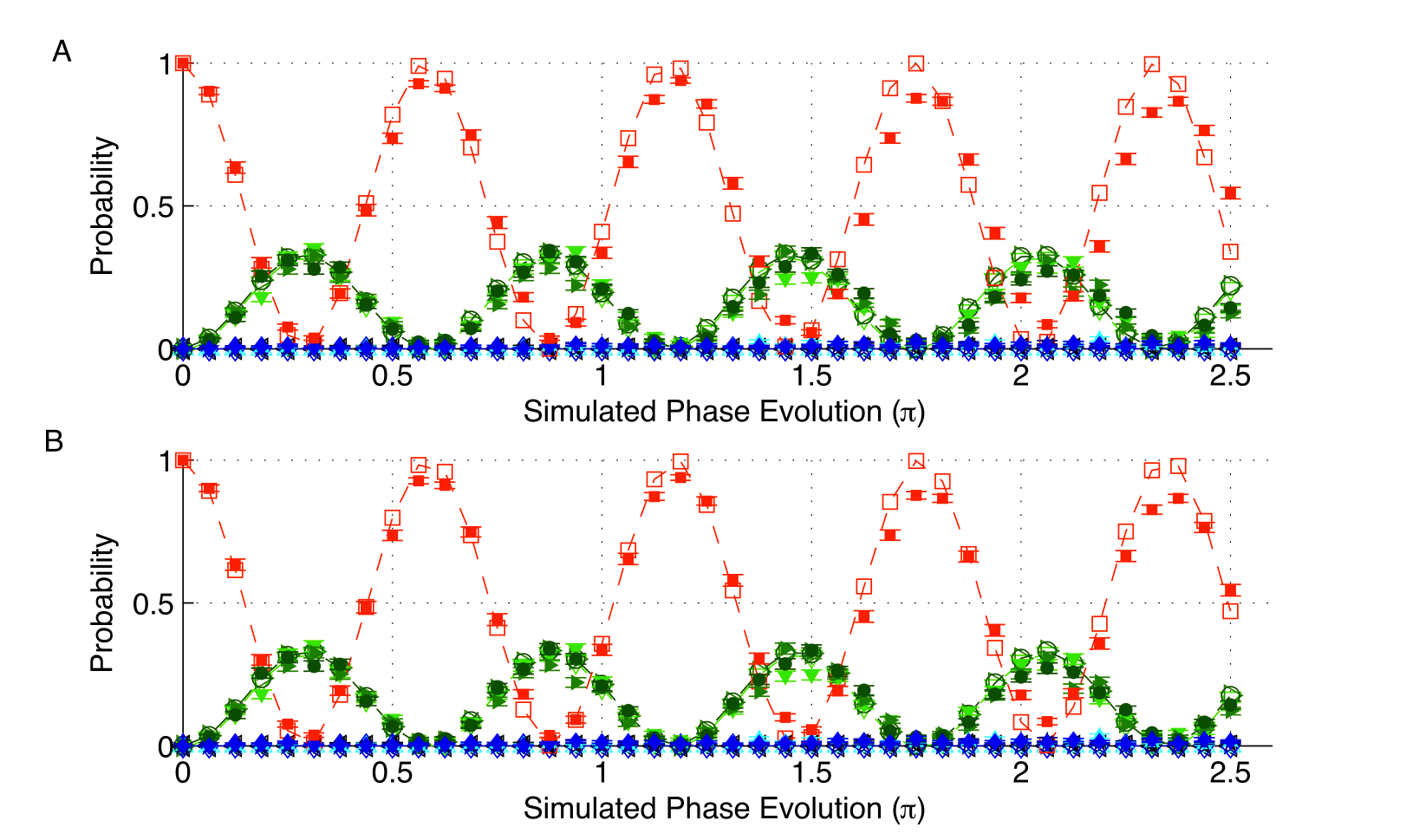}
\end{center}
\vspace{0mm}
\caption{
\textbf{Sensitivity of the simulations to imperfectly set operations.}
Simulation results of a three-spin Ising model (shown in Fig. 3A, main text) are compared with theoretical predictions for the cases of \textbf{(A)} ideal gate operations \textbf{(B)} non-ideal gate operations: the phase $\theta$ of the $O_4$ operation used to simulate the spin-spin interaction in each digital step is wrong by 1\%. The frequency mismatch in the previous panel is now largely corrected.
}
\label{S9}
\end{figure}

The 3-spin transverse Ising model results, presented in Figure 3A in the main text, exhibit a slight frequency shift compared to the ideal digitised case. These effects could easily be the result of errors made in the setting-up/optimising of the gate operations required for the sequence. Figure~\ref{S9} shows how the observed frequency mismatch would be expected if the phase angle $\theta$ of the $O_4$ operation used in each trotter step is set incorrectly by only $1\%$. The sensitivity to these effects suggests the need for future work on developing even more accurate methods to optimise gate operations in the lab than are currently employed~\cite{Benhelm:2008b}.

\clearpage

\begin{table}\footnotesize
\caption{Measured output state fidelities after the 3-qubit operation $U(\pi/4){=}exp({-i\sigma_z\sigma_x\sigma_x\pi/4})$,  for the orthogonal basis set of input states shown. Ideally these are eigenstates and should be unchanged by the operation. The average output state fidelity is $F_1{=}0.942(5)$.}
\begin{center}
\begin{tabular}{c||c}
\hline
Input & Fidelity\\
\hline
$|0\rangle|++\rangle_x$  & 0.94(1)\\
$|0\rangle|+-\rangle_x$ & 0.94(2)\\
$|0\rangle|-+\rangle_x$ & 0.97(1)\\
$|0\rangle|--\rangle_x$ & 0.95(1)\\
$|1\rangle|++\rangle_x$ & 0.93(2)\\
$|1\rangle|+-\rangle_x$ & 0.95(1)\\
$|1\rangle|-+\rangle_x$ & 0.93(2)\\
$|1\rangle|--\rangle_x$ & 0.93(2)\\
\hline
\end{tabular}
\end{center}
\label{T1}
\end{table}%

\begin{table}\footnotesize
\caption{Measured output state fidelities after the 3-qubit operation $U(\pi/4){=}exp({-i\sigma_z\sigma_x\sigma_x\pi/4})$,  for the orthogonal basis set of input states shown. Ideally these should become entangled states. Fidelities are derived from 3 parity measurements and two logical populations (extracted from one measurement in the logical basis). The average output state fidelity is $F_2{=}0.908(6)$. Ideally the populations should each be 0.5 and the absolute value of each parity should be 1.}
\begin{center}
\begin{tabular}{c||ccc|cc||c}
\hline
Input & \multicolumn{3}{c}{Parity}\vline& \multicolumn{2}{c}{Populations}\vline & Fidelity\\\
& 1&2& 3&1&2&\\
\hline

$|+++\rangle_y$   & 0.88(4) & -0.89(4) & 0.86(4) & 0.48(3) & 0.48(3)& 0.92(2)\\
$|++-\rangle_y$ & 0.91(4) & -0.87(4) & 0.89(4) & 0.56(3) & 0.35(3)& 0.90(2) \\
$|+-+\rangle_y$ & 0.84(4) & -0.94(4) & 0.92(4) & 0.40(3) & 0.45(3) & 0.87(2)\\
$|-++\rangle_y$ & 0.86(4) & -0.86(4) & 0.94(4) & 0.46(3) & 0.45(3)& 0.90(2) \\
$|+--\rangle_y$ & 0.92(4) & -0.87(4) & 0.94(4) & 0.51(3) & 0.41(3)& 0.91(2)  \\
$|-+-\rangle_y$ & 0.90(4) & -0.92(4) & 0.94(4) & 0.44(3) & 0.48(3) & 0.92(2)\\
$|--+\rangle_y$ & 0.94(4) & -0.83(4) & 0.95(4) & 0.44(3) & 0.48(3)& 0.91(2) \\
$|---\rangle_y$ & 0.88(4) & -0.86(4) & 0.91(4) & 0.47(3) & 0.49(3) & 0.92(2)\\
\hline
\end{tabular}
\end{center}
\label{T2}
\end{table}%

\begin{table}\footnotesize
\caption{Measured output state fidelities after the 6-qubit operation $U(\pi/4){=}exp({-i\sigma_y\sigma_x\sigma_x\sigma_x\sigma_x\sigma_x\pi/4})$,  for the orthogonal set of input states shown. Ideally these are eigenstates and should be unchanged by the operation. This table shows 32 of the 64 states that form a complete basis. The other 32 are shown in Table IV. The average output state fidelity between both tables is $F_1{=}0.792(4)$.}
\begin{center}
\begin{tabular}{c||c}
\hline
Input & Fidelity\\
\hline
$|{+}\rangle_y|{{+}}{{+}}{{+}}{{+}}{{+}}\rangle_x$    &0.78(3)\\
$|{+}\rangle_y|{+}{+}{+}{+}{-}\rangle_x$    &0.81(3)\\
$|{+}\rangle_y|{+}{+}{+}{-}{+}\rangle_x$    &0.83(3)\\
$|{+}\rangle_y|{+}{+}{+}{-}{-}\rangle_x$    &0.77(3)\\
$|{+}\rangle_y|{+}{+}{-}{+}{+}\rangle_x$   &0.78(3)\\
$|{+}\rangle_y|{+}{+}{-}{+}{-}\rangle_x$   &0.75(3)\\
$|{+}\rangle_y|{+}{+}{-}{-}{+}\rangle_x$    &0.75(3)\\
$|{+}\rangle_y|{+}{+}{-}{-}{-}\rangle_x$    &0.80(3)\\
$|{+}\rangle_y|{+}{-}{+}{+}{+}\rangle_x$    &0.76(3)\\
$|{+}\rangle_y|{+}{-}{+}{+}{-}\rangle_x$    &0.79(3)\\
$|{+}\rangle_y|{+}{-}{+}{-}{+}\rangle_x$    &0.75(3)\\
$|{+}\rangle_y|{+}{-}{+}{-}{-}\rangle_x$   &0.71(3)\\
$|{+}\rangle_y|{+}{-}{-}{+}{+}\rangle_x$    &0.74(3)\\
$|{+}\rangle_y|{+}{-}{-}{+}{-}\rangle_x$    &0.77(3)\\
$|{+}\rangle_y|{+}{-}{-}{-}{+}\rangle_x$    &0.87(2)\\
$|{+}\rangle_y|{+}{-}{-}{-}{-}\rangle_x$    &0.79(3)\\
$|{+}\rangle_y|{-}{+}{+}{+}{+}\rangle_x$    &0.77(3)\\
$|{+}\rangle_y|{-}{+}{+}{+}{-}\rangle_x$    &0.86(2)\\
$|{+}\rangle_y|{-}{+}{+}{-}{+}\rangle_x$    &0.79(3)\\
$|{+}\rangle_y|{-}{+}{+}{-}{-}\rangle_x$    &0.73(3)\\
$|{+}\rangle_y|{-}{+}{-}{+}{+}\rangle_x$    &0.78(3)\\
$|{+}\rangle_y|{-}{+}{-}{+}{-}\rangle_x$    &0.77(3)\\
$|{+}\rangle_y|{-}{+}{-}{-}{+}\rangle_x$    &0.83(3)\\
$|{+}\rangle_y|{-}{+}{-}{-}{-}\rangle_x$    &0.77(3)\\
$|{+}\rangle_y|{-}{-}{+}{+}{+}\rangle_x$    &0.76(3)\\
$|{+}\rangle_y|{-}{-}{+}{+}{-}\rangle_x$   &0.75(3)\\
$|{+}\rangle_y|{-}{-}{+}{-}{+}\rangle_x$    &0.89(2)\\
$|{+}\rangle_y|{-}{-}{+}{-}{-}\rangle_x$   &0.79(3)\\
$|{+}\rangle_y|{-}{-}{-}{+}{+}\rangle_x$    &0.83(3)\\
$|{+}\rangle_y|{-}{-}{-}{+}{-}\rangle_x$    &0.81(3)\\
$|{+}\rangle_y|{-}{-}{-}{-}{+}\rangle_x$    &0.82(3)\\
$|{+}\rangle_y|{-}{-}{-}{-}{-}\rangle_x$    &0.80(3)\\
\hline
\end{tabular}
\end{center}
\label{T3}
\end{table}%

\begin{table}\footnotesize
\caption{Measured output state fidelities after the 6-qubit operation $U(\pi/4){=}exp({-i\sigma_y\sigma_x\sigma_x\sigma_x\sigma_x\sigma_x\pi/4})$,  for the orthogonal set of input states shown. Ideally these are eigenstates and should be unchanged by the operation. This table shows 32 of the 64 states that form a complete basis. The other 32 are shown in Table III. The average output state fidelity between both tables is $F_1{=}0.792(4)$.}
\begin{center}
\begin{tabular}{c||c}
\hline
Input & Fidelity\\
\hline

$|{-}\rangle_y|{+}{+}{+}{+}{+}\rangle_x$    &0.81(3)\\
$|{-}\rangle_y|{+}{+}{+}{+}{-}\rangle_x$    &0.78(3)\\
$|{-}\rangle_y|{+}{+}{+}{-}{+}\rangle_x$    &0.79(3)\\
$|{-}\rangle_y|{+}{+}{+}{-}{-}\rangle_x$    &0.82(3)\\
$|{-}\rangle_y|{+}{+}{-}{+}{+}\rangle_x$    &0.77(3)\\
$|{-}\rangle_y|{+}{+}{-}{+}{-}\rangle_x$    &0.80(3)\\
$|{-}\rangle_y|{+}{+}{-}{-}{+}\rangle_x$    &0.85(3)\\
$|{-}\rangle_y|{+}{+}{-}{-}{-}\rangle_x$    &0.81(3)\\
$|{-}\rangle_y|{+}{-}{+}{+}{+}\rangle_x$    &0.84(3)\\
$|{-}\rangle_y|{+}{-}{+}{+}{-}\rangle_x$    &0.79(3)\\
$|{-}\rangle_y|{+}{-}{+}{-}{+}\rangle_x$    &0.82(3)\\
$|{-}\rangle_y|{+}{-}{+}{-}{-}\rangle_x$    &0.78(3)\\
$|{-}\rangle_y|{+}{-}{-}{+}{+}\rangle_x$    &0.77(3)\\
$|{-}\rangle_y|{+}{-}{-}{+}{-}\rangle_x$    &0.71(3)\\
$|{-}\rangle_y|{+}{-}{-}{-}{+}\rangle_x$    &0.77(3)\\
$|{-}\rangle_y|{+}{-}{-}{-}{-}\rangle_x$    &0.83(3)\\
$|{-}\rangle_y|{-}{+}{+}{+}{+}\rangle_x$    &0.80(3)\\
$|{-}\rangle_y|{-}{+}{+}{+}{-}\rangle_x$    &0.80(3)\\
$|{-}\rangle_y|{-}{+}{+}{-}{+}\rangle_x$    &0.83(3)\\
$|{-}\rangle_y|{-}{+}{+}{-}{-}\rangle_x$    &0.78(3)\\
$|{-}\rangle_y|{-}{+}{-}{+}{+}\rangle_x$    &0.78(3)\\
$|{-}\rangle_y|{-}{+}{-}{+}{-}\rangle_x$    &0.72(3)\\
$|{-}\rangle_y|{-}{+}{-}{-}{+}\rangle_x$    &0.83(3)\\
$|{-}\rangle_y|{-}{+}{-}{-}{-}\rangle_x$    &0.83(3)\\
$|{-}\rangle_y|{-}{-}{+}{+}{+}\rangle_x$    &0.78(3)\\
$|{-}\rangle_y|{-}{-}{+}{+}{-}\rangle_x$    &0.80(3)\\
$|{-}\rangle_y|{-}{-}{+}{-}{+}\rangle_x$    &0.83(3)\\
$|{-}\rangle_y|{-}{-}{+}{-}{-}\rangle_x$    &0.83(3)\\
$|{-}\rangle_y|{-}{-}{-}{+}{+}\rangle_x$    &0.80(3)\\
$|{-}\rangle_y|{-}{-}{-}{+}{-}\rangle_x$    &0.85(3)\\
$|{-}\rangle_y|{-}{-}{-}{-}{+}\rangle_x$   &0.81(3)\\
$|{-}\rangle_y|{-}{-}{-}{-}{-}\rangle_x$   &0.81(3)\\
\hline
\end{tabular}
\end{center}
\label{T4}
\end{table}

\begin{table}\footnotesize
\caption{Measured output state fidelities after the 6-qubit operation $U(\pi/4){=}exp({-i\sigma_y\sigma_x\sigma_x\sigma_x\sigma_x\sigma_x\pi/4})$,  for the orthogonal set of input states shown. Ideally these should become entangled states. Fidelities are derived from 6 parity measurements and two logical populations (extracted from one measurement in the logical basis). This table shows 32 of the 64 states that form a complete basis. The other 32 are shown in Table VI. The average output state fidelity between both tables is $F_1{=}0.767(6)$. Ideally the populations should each be 0.5 and the absolute value of each parity should be 1.}
\begin{center}
\begin{tabular}{c|cccccc|cc|c}
\hline
Input & \multicolumn{6}{c}{Parity}\vline& \multicolumn{2}{c}{Populations}\vline & Fidelity\\
& 1&2& 3&4&5&6&1&2&\\
\hline
$|000000\rangle$&    0.64(8)&  -0.64(8)&   0.62(8)&  -0.62(8)&   0.64(8)&  -0.66(8)&   0.34(5)&   0.45(5)&   0.71(5)\\
$|000001\rangle$&    0.74(7)&  -0.64(8)&   0.80(6)&  -0.58(8)&   0.86(5)&  -0.66(8)&   0.33(5)&   0.51(5)&   0.78(5)\\
$|000010\rangle$&    0.76(6)&  -0.76(6)&   0.74(7)&  -0.72(7)&   0.74(7)&  -0.76(6)&   0.46(5)&   0.38(5)&   0.79(5)\\
$|000011\rangle$&    0.84(5)&  -0.74(7)&   0.66(8)&  -0.78(6)&   0.76(6)&  -0.80(6)&   0.35(5)&   0.46(5)&   0.79(5)\\
$|000100\rangle$&    0.72(7)&  -0.66(8)&   0.68(7)&  -0.68(7)&   0.72(7)&  -0.76(6)&   0.32(5)&   0.46(5)&   0.74(5)\\
$|000101\rangle$&   0.68(7)&  -0.72(7)&   0.86(5)&  -0.86(5)&   0.72(7)&  -0.76(6)&   0.41(5)&   0.37(5)&   0.77(5)\\
$|000110\rangle$&    0.84(5)&  -0.80(6)&   0.86(5)&  -0.80(6)&   0.74(7)&  -0.88(5)&   0.43(5)&   0.38(5)&   0.82(5)\\
$|000111\rangle$&    0.88(5)&  -0.68(7)&   0.72(7)&  -0.82(6)&   0.70(7)&  -0.82(6)&   0.35(5)&   0.46(5)&   0.79(5)\\
$|001000\rangle$&    0.80(6)&  -0.64(8)&   0.54(8)&  -0.76(6)&   0.70(7)&  -0.72(7)&   0.46(5)&   0.39(5)&   0.77(5)\\
$|001001\rangle$&    0.84(5)&  -0.66(8)&   0.68(7)&  -0.68(7)&   0.82(6)&  -0.76(6)&   0.37(5)&   0.42(5)&   0.76(5)\\
$|001010\rangle$&    0.84(5)&  -0.72(7)&   0.72(7)&  -0.80(6)&   0.62(8)&  -0.74(7)&   0.43(5)&   0.39(5)&   0.78(5)\\
$|001011\rangle$&    0.84(5)&  -0.58(8)&   0.68(7)&  -0.80(6)&   0.70(7)&  -0.78(6)&   0.37(5)&   0.48(5)&   0.79(5)\\
$|001100\rangle$&    0.80(6)&  -0.72(7)&   0.76(6)&  -0.82(6)&   0.64(8)&  -0.86(5)&   0.46(5)&   0.37(5)&   0.80(5)\\
$|001101\rangle$&    0.74(7)&  -0.76(6)&   0.86(5)&  -0.72(7)&   0.70(7)&  -0.86(5)&   0.39(5)&   0.44(5)&   0.80(5)\\
$|001110\rangle$&    0.82(6)&  -0.74(7)&   0.86(5)&  -0.80(6)&   0.76(6)&  -0.82(6)&   0.51(5)&   0.28(4)&   0.80(5)\\
$|001111\rangle$&    0.88(5)&  -0.64(8)&   0.52(9)&  -0.74(7)&   0.72(7)&  -0.76(6)&   0.28(4)&   0.55(5)&   0.77(5)\\
$|010000\rangle$&    0.88(5)&  -0.72(7)&   0.74(7)&  -0.74(7)&   0.74(7)&  -0.84(5)&   0.44(5)&   0.38(5)&   0.80(5)\\
$|010001\rangle$&    0.76(6)&  -0.74(7)&   0.66(8)&  -0.80(6)&   0.78(6)&  -0.84(5)&   0.36(5)&   0.41(5)&   0.77(5)\\
$|010010\rangle$&    0.84(5)&  -0.68(7)&   0.88(5)&  -0.76(6)&   0.72(7)&  -0.60(8)&   0.40(5)&   0.48(5)&   0.81(5)\\
$|010011\rangle$&    0.76(6)&  -0.64(8)&   0.78(6)&  -0.62(8)&   0.80(6)&  -0.76(6)&   0.36(5)&   0.46(5)&   0.77(5)\\
$|010100\rangle$&    0.88(5)&  -0.78(6)&   0.76(6)&  -0.82(6)&   0.78(6)&  -0.80(6)&   0.45(5)&   0.45(5)&   0.85(5)\\
$|010101\rangle$&    0.78(6)&  -0.84(5)&   0.72(7)&  -0.78(6)&   0.76(6)&  -0.86(5)&   0.28(4)&   0.46(5)&   0.77(5)\\
$|010110\rangle$&    0.86(5)&  -0.86(5)&   0.82(6)&  -0.92(4)&   0.84(5)&  -0.82(6)&   0.45(5)&   0.42(5)&   0.86(5)\\
$|010111\rangle$&    0.82(6)&  -0.80(6)&   0.78(6)&  -0.72(7)&   0.72(7)&  -0.76(6)&   0.23(4)&   0.56(5)&   0.78(5)\\
$|011000\rangle$&    0.76(6)&  -0.72(7)&   0.80(6)&  -0.78(6)&   0.74(7)&  -0.86(5)&   0.38(5)&   0.50(5)&   0.83(5)\\
$|011001\rangle$&    0.86(5)&  -0.72(7)&   0.78(6)&  -0.76(6)&   0.72(7)&  -0.80(6)&   0.33(5)&   0.43(5)&   0.77(5)\\
$|011010\rangle$&    0.80(6)&  -0.74(7)&   0.68(7)&  -0.76(6)&   0.78(6)&  -0.90(4)&   0.49(5)&   0.37(5)&   0.82(5)\\
$|011011\rangle$&    0.78(6)&  -0.80(6)&   0.56(8)&  -0.56(8)&   0.66(8)&  -0.70(7)&   0.29(5)&   0.50(5)&   0.73(5)\\
$|011100\rangle$&    0.84(5)&  -0.82(6)&   0.76(6)&  -0.68(7)&   0.72(7)&  -0.78(6)&   0.48(5)&   0.37(5)&   0.81(5)\\
$|011101\rangle$&    0.82(6)&  -0.76(6)&   0.46(9)&  -0.42(9)&   0.40(9)&  -0.64(8)&   0.31(5)&   0.45(5)&   0.67(5)\\
$|011110\rangle$&    0.90(4)&  -0.68(7)&   0.76(6)&  -0.70(7)&   0.66(8)&  -0.80(6)&   0.50(5)&   0.33(5)&   0.79(5)\\
$|011111\rangle$&    0.64(8)&  -0.58(8)&   0.64(8)&  -0.68(7)&   0.70(7)&  -0.74(7)&   0.26(4)&   0.57(5)&   0.75(5)\\
\hline
\end{tabular}
\end{center}
\label{T5}
\end{table}%

\begin{table}\footnotesize
\caption{Measured output state fidelities after the 6-qubit operation $U(\pi/4){=}exp({-i\sigma_y\sigma_x\sigma_x\sigma_x\sigma_x\sigma_x\pi/4})$,  for the orthogonal set of input states shown. Ideally these should become entangled states. Fidelities are derived from 6 parity measurements and two logical populations (extracted from one measurement in the logical basis). This table shows 32 of the 64 states that form a complete basis. The other 32 are shown in Table V. The average output state fidelity between both tables is $F_1{=}0.767(6)$. Ideally the populations should each be 0.5 and the absolute value of each parity should be 1.}
\begin{center}
\begin{tabular}{c|cccccc|cc|c}
\hline
Input & \multicolumn{6}{c}{Parity}\vline& \multicolumn{2}{c}{Populations}\vline & Fidelity\\
& 1&2& 3&4&5&6&1&2&\\
\hline
$|100000\rangle$&    0.70(7)&  -0.68(7)&   0.86(5)&  -0.70(7)&   0.72(7)&  -0.80(6)&   0.33(5)&   0.53(5)&   0.80(5)\\
$|100001\rangle$&    0.64(8)&  -0.78(6)&   0.80(6)&  -0.80(6)&   0.74(7)&  -0.60(8)&   0.55(5)&   0.26(4)&   0.77(5)\\
$|100010\rangle$&    0.86(5)&  -0.78(6)&   0.86(5)&  -0.74(7)&   0.84(5)&  -0.70(7)&   0.37(5)&   0.51(5)&   0.84(5)\\
$|100011\rangle$&    0.78(6)&  -0.70(7)&   0.70(7)&  -0.72(7)&   0.70(7)&  -0.74(7)&   0.41(5)&   0.43(5)&   0.78(5)\\
$|100100\rangle$&    0.86(5)&  -0.76(6)&   0.72(7)&  -0.78(6)&   0.72(7)&  -0.82(6)&   0.45(5)&   0.40(5)&   0.81(5)\\
$|100101\rangle$&    0.80(6)&  -0.66(8)&   0.66(8)&  -0.80(6)&   0.86(5)&  -0.80(6)&   0.44(5)&   0.37(5)&   0.79(5)\\
$|100110\rangle$&    0.86(5)&  -0.74(7)&   0.78(6)&  -0.74(7)&   0.80(6)&  -0.80(6)&   0.34(5)&   0.47(5)&   0.80(5)\\
$|100111\rangle$&    0.84(5)&  -0.76(6)&   0.70(7)&  -0.54(8)&   0.64(8)&  -0.76(6)&   0.45(5)&   0.35(5)&   0.75(5)\\
$|101000\rangle$&    0.80(6)&  -0.74(7)&   0.78(6)&  -0.76(6)&   0.80(6)&  -0.86(5)&   0.40(5)&   0.46(5)&   0.83(5)\\
$|101001\rangle$&    0.72(7)&  -0.90(4)&   0.64(8)&  -0.72(7)&   0.78(6)&  -0.84(5)&   0.35(5)&   0.46(5)&   0.79(5)\\
$|101010\rangle$&    0.86(5)&  -0.72(7)&   0.74(7)&  -0.76(6)&   0.88(5)&  -0.76(6)&   0.41(5)&   0.47(5)&   0.83(5)\\
$|101011\rangle$&    0.78(6)&  -0.62(8)&   0.36(9)&  -0.54(8)&   0.66(8)&  -0.82(6)&   0.45(5)&   0.38(5)&   0.73(5)\\
$|101100\rangle$&    0.90(4)&  -0.82(6)&   0.68(7)&  -0.52(9)&   0.74(7)&  -0.82(6)&   0.36(5)&   0.46(5)&   0.78(5)\\
$|101101\rangle$&    0.82(6)&  -0.68(7)&   0.46(9)&  -0.60(8)&   0.52(9)&  -0.86(5)&   0.40(5)&   0.33(5)&   0.69(5)\\
$|101110\rangle$&    0.80(6)&  -0.56(8)&   0.56(8)&  -0.72(7)&   0.66(8)&  -0.60(8)&   0.51(5)&   0.35(5)&   0.76(5)\\
$|101111\rangle$&    0.68(7)&  -0.70(7)&   0.48(9)&  -0.60(8)&   0.66(8)&  -0.78(6)&   0.35(5)&   0.46(5)&   0.73(5)\\
$|110000\rangle$&    0.74(7)&  -0.78(6)&   0.52(9)&  -0.76(6)&   0.72(7)&  -0.76(6)&   0.37(5)&   0.49(5)&   0.79(5)\\
$|110001\rangle$&   0.76(6)&  -0.88(5)&   0.72(7)&  -0.72(7)&   0.80(6)&  -0.70(7)&   0.50(5)&   0.34(5)&   0.80(5)\\
$|110010\rangle$&    0.78(6)&  -0.72(7)&   0.68(7)&  -0.62(8)&   0.68(7)&  -0.80(6)&   0.42(5)&   0.44(5)&   0.79(5)\\
$|110011\rangle$&    0.86(5)&  -0.58(8)&   0.62(8)&  -0.54(8)&   0.62(8)&  -0.80(6)&   0.37(5)&   0.39(5)&   0.72(5)\\
$|110100\rangle$&    0.88(5)&  -0.74(7)&   0.82(6)&  -0.70(7)&   0.64(8)&  -0.82(6)&   0.36(5)&   0.48(5)&   0.80(5)\\
$|110101\rangle$&    0.88(5)&  -0.48(9)&   0.48(9)&  -0.48(9)&   0.48(9)&  -0.62(8)&   0.35(5)&   0.29(5)&   0.61(5)\\
$|110110\rangle$&    0.82(6)&  -0.54(8)&   0.48(9)&  -0.64(8)&   0.64(8)&  -0.82(6)&   0.48(5)&   0.37(5)&   0.75(5)\\
$|110111\rangle$&    0.82(6)&  -0.78(6)&   0.54(8)&  -0.3(1)  &   0.74(7)&  -0.72(7)&   0.43(5)&   0.44(5)&   0.76(5)\\
$|111000\rangle$&    0.72(7)&  -0.80(6)&   0.72(7)  &  -0.54(8)&   0.72(7)&  -0.78(6)&   0.47(5)&   0.38(5)&   0.78(5)\\
$|111001\rangle$&    0.70(7)&  -0.50(9)&   0.60(8)&  -0.3(1)  &   0.52(9)&  -0.80(6)&   0.34(5)&   0.44(5)&   0.67(5)\\
$|111010\rangle$&    0.70(7)&  -0.68(7)&   0.1(1)  &  -0.54(8)&   0.62(8)&  -0.78(6)&   0.48(5)&   0.38(5)&   0.72(5)\\
$|111011\rangle$&    0.66(8)&  -0.34(9)&   0.52(9)&  -0.38(9)&   0.36(9)&  -0.62(8)&   0.50(5)&   0.36(5)&   0.67(5)\\
$|111100\rangle$&    0.72(7)&  -0.80(6)&   0.2(1)   & -0.40(9)&   0.54(8)&  -0.74(7)&   0.40(5)&   0.32(5)&   0.64(5)\\
$|111101\rangle$&    0.80(6)&  -0.56(8)&   0.38(9)&  -0.3(1)  &   0.56(8)&  -0.54(8)&   0.40(5)&   0.41(5)&   0.67(5)\\
$|111110\rangle$&    0.80(6)&  -0.60(8)&   0.54(8)&  -0.76(6)&   0.76(6)&  -0.68(7)&   0.48(5)&   0.32(5)&   0.75(5)\\
$|111111\rangle$&    0.40(9)&  -0.72(7)&   0.32(9)&  -0.68(7)&   0.34(9)&  -0.68(7)&   0.38(5)&   0.39(5)&   0.65(5)\\
\hline
\end{tabular}
\end{center}
\label{T6}
\end{table}%

\clearpage{}

\bibliographystyle{Science}
\bibliography{trotterbib}

\end{document}